\documentclass[11pt,preprint]{aastex}

\begin{document}

\shorttitle{Dust Disks in $\rho$ Oph}

\shortauthors{Andrews \& Williams}

\title{A Submillimeter View of Circumstellar Dust Disks in $\rho$ Ophiuchus}

\author{Sean M. Andrews\altaffilmark{1} and Jonathan P. Williams}

\affil{Institute for Astronomy, University of Hawaii, 2680 Woodlawn Drive, Honolulu, HI 96822}
\email{andrews@ifa.hawaii.edu, jpw@ifa.hawaii.edu}

\altaffiltext{1}{Present address: Harvard-Smithsonian Center for Astrophysics, 
60 Garden Street, MS 42, Cambridge, MA 02138}

\begin{abstract}
We present new multiwavelength submillimeter continuum measurements of the 
circumstellar dust around 48 young stars in the $\rho$ Ophiuchus dark clouds.  
Supplemented with previous 1.3\,mm observations of an additional 99 objects 
from the literature, the statistical distributions of disk masses and 
submillimeter colors are calculated and compared to those in the Taurus-Auriga 
region.  These basic submillimeter properties of young stellar objects in both 
environments are shown to be essentially identical.  As with their Taurus 
counterparts, the $\rho$ Oph circumstellar dust properties are shown to evolve 
along an empirical evolution sequence based on the infrared spectral energy 
distribution.  The combined $\rho$ Oph and Taurus Class II samples (173 
sources) are used to set benchmark values for basic outer disk characteristics: 
$M_d \sim 0.005$\,M$_{\odot}$, $M_d/M_{\ast} \sim 1$\%, and $\alpha \sim 2$ 
(where $F_{\nu} \propto \nu^{\alpha}$ between 350\,$\mu$m and 1.3\,mm).  The 
precision of these numbers are addressed in the context of substantial solid 
particle growth in the earliest stages of the planet formation process.  There 
is some circumstantial evidence that disk masses inferred from submillimeter 
emission may be under-estimated by up to an order of magnitude.
\end{abstract}

\section{Introduction}

Circumstellar accretion disks are an integral part of the star and planet 
formation process.  They are critical for conserving angular momentum during 
molecular cloud core collapse \citep{terebey84}, channeling mass onto the 
central star \citep{lyndenbell74}, and assembling raw material into a planetary 
system \citep[e.g.,][]{lissauer07}.  The structural, material, and chemical 
compositions of disks provide crucial insights on the initial conditions of the 
planet formation process.  While molecular gas is the dominant mass species in 
these disks, observations of this phase are notoriously difficult due to the 
radiative inefficiency of H$_2$, the depletion of molecules from the gas phase 
onto dust grain surfaces, and the relative insensitivity of many current 
telescopes.  Consequently, much of our knowledge of disk properties is gathered 
from the comparatively bright thermal continuum emission of dust particles 
heated by irradiation from the central star \citep[see][]{natta07}.  

At any given frequency, this continuum emission is generated from dust grains 
at a relatively small range of contributing stellocentric radii 
\citep[e.g.,][]{beckwith90}.  Therefore, the detailed morphology of its 
spectral energy distribution (SED) is dictated by the physical structure and 
material content of the disk (e.g., temperatures, densities, geometry, and 
particle properties).  Much of the infrared SED is generated in the optically 
thick inner few AU of the disk, corresponding roughly to the region in the 
Solar System occupied by the terrestrial planets.  Because optical depths are 
so high at these radii, the infrared SED can be used to infer disk temperatures 
\citep[e.g.,][]{adams87,chiang97,dalessio98}.  The far-infrared and 
millimeter/radio portion of the SED (hereafter referred to as ``submillimeter" 
for convenience) originates primarily in the optically thin outer disk, from 
several to perhaps hundreds of AU.  The low column densities in this region 
lead to a direct relationship between the submillimeter luminosity and the 
disk mass \citep{hildebrand83,beckwith90,am94}.  Moreover, the shape of the 
long-wavelength SED is at least partially controlled by the disk opacity 
\citep{beckwith91,mannings94}.  The optical properties of the dust grain 
population, especially the particle size distribution, are encoded in this 
latter parameter \citep{pollack94,draine06}.  

In practice, one can infer the circumstellar dust mass around a young star with 
a submillimeter flux density and constrain the bulk optical characteristics of 
the solid particle population using submillimeter colors.  Constraints on these 
properties are empirical prerequisites for understanding the mechanisms and 
timescales involved in the planet formation process.  The progenitor disk mass 
not only sets a limit on the material composition and structure of a potential 
planetary system, but can also elucidate whether any particular mechanism of 
planet formation is more plausible than another.  As an example, models that 
rely on gravitational instabilities in the disk to drive rapid giant planet 
formation are only tenable if the disk is sufficiently massive 
\citep{boss98,mayer07}.  Moreover, interpreting submillimeter colors in terms 
of the current solid particle size distribution in these disks may reveal clues 
on the efficiency of the grain growth process, representing the first tentative 
steps of building planetesimals.  

\citet{beckwith90} presented the first large survey of the 1.3\,mm emission 
from young stellar objects (YSOs; defined here as a young star and its 
circumstellar environment) in the Taurus-Auriga star formation region.  
Following up with a supplementary survey \citep{osterloh95} and smaller,   
multiwavelength submillimeter samples \citep{beckwith91,mannings94}, these 
studies provided the first detailed look at the circumstellar disk mass 
distribution and the shape of the opacity spectrum.  The results argued for a 
typical disk mass similar to that required to form the Solar System and an 
opacity spectrum shallower than noted in the interstellar medium (ISM), 
possibly due to grain growth.  Around the same time, \citet{am94} built on 
their previous work \citep{andre90} with a pioneering investigation of the 
millimeter continuum emission from YSOs in the $\rho$ Ophiuchus dark clouds.  
One major impact of their work was highlighting that the millimeter 
luminosities appeared tied to the same YSO evolution sequence typically tracked 
in the infrared: younger sources on this empirical sequence were shown to have 
larger circumstellar dust masses.  

To build on these landmark studies while taking advantage of technological 
advancements in the sensitivity and efficiency of submillimeter detectors, we 
undertook a comprehensive, deep, multiwavelength survey of Taurus YSOs 
\citep[][hereafter Paper I]{aw05}.  Centered around an 850\,$\mu$m sample of 
$\sim$150 sources, this work had the benefits of at least a factor of 5 better 
sensitivity and a considerably more uniform luminosity limit than previous 
observations.  In addition to a more complete census of disk masses and their 
evolution, this survey permitted us to derive the distribution of submillimeter 
colors between 350\,$\mu$m and 1.3\,mm, and show that these also changed as a 
function of YSO evolutionary state.  In this article, we present a similar 
study of the submillimeter continuum emission from circumstellar dust in the 
$\rho$ Oph region.  In \S 2 we describe the new survey data, its acquisition 
and reduction, and the compilation of an extended sample similar to the one 
used for Taurus.  Disk masses and colors are computed in \S 3, and the results 
for $\rho$ Oph YSOs are systematically compared with their counterparts in 
Taurus.  Potential relationships of these properties with evolutionary 
diagnostics, stellar host parameters, and the planet formation process are 
discussed in \S 4, before concluding with a summary in \S 5.

\section{Observations and Data Reduction}

The Submillimeter Common-User Bolometer Array \citep[SCUBA:][]{holland99} 
instrument at the 15\,m James Clerk Maxwell Telescope (JCMT) was used to 
observe 48 $\rho$ Oph YSOs during two runs in 2004 April and 2005 June.  The 
observations were conducted in photometry mode, providing simultaneous 
measurements of the 450 and 850\,$\mu$m continuum emission within FWHM beam 
diameters of 9\arcsec\ and 14\arcsec, respectively.  Pointing centers accurate 
to $< 1$\%\ of the beam size were determined from the Two Micron All Sky Survey 
(2MASS) Point Source Catalog \citep{cutri03}.  The typical precipitable water 
vapor level during the observations was $\lesssim 1.8$\,mm, corresponding to 
zenith optical depths of 0.3$-$0.4 at 850\,$\mu$m and $\sim$2 at 450\,$\mu$m.  
The same observing and data reduction strategies described in detail in Paper I 
were utilized for this sample.  The systematic uncertainties associated with 
absolute flux calibration are expected to be $\sim$10\%\ and 25\%\ for 850 and 
450\,$\mu$m, respectively.  

Images of the 350\,$\mu$m continuum for 7 YSOs were obtained with the 
Submillimeter High Angular Resolution Camera \citep[SHARC-II:][]{dowell03} on 
the 10\,m Caltech Submillimeter Observatory (CSO) telescope in 2004 March and 
June.  As with the SHARC-II observations described in Paper I, these 
diffraction-limited images (FWHM beam diameter of 9\arcsec) were obtained by 
constantly scanning the telescope in the vicinity of the source in a Lissajous 
pattern.  Further details on the observing technique and data reduction are 
highlighted in Paper I.  The absolute flux calibration of these 350\,$\mu$m 
data is accurate to $\sim$25\%.   

\begin{figure}
\epsscale{1.0}
\plotone{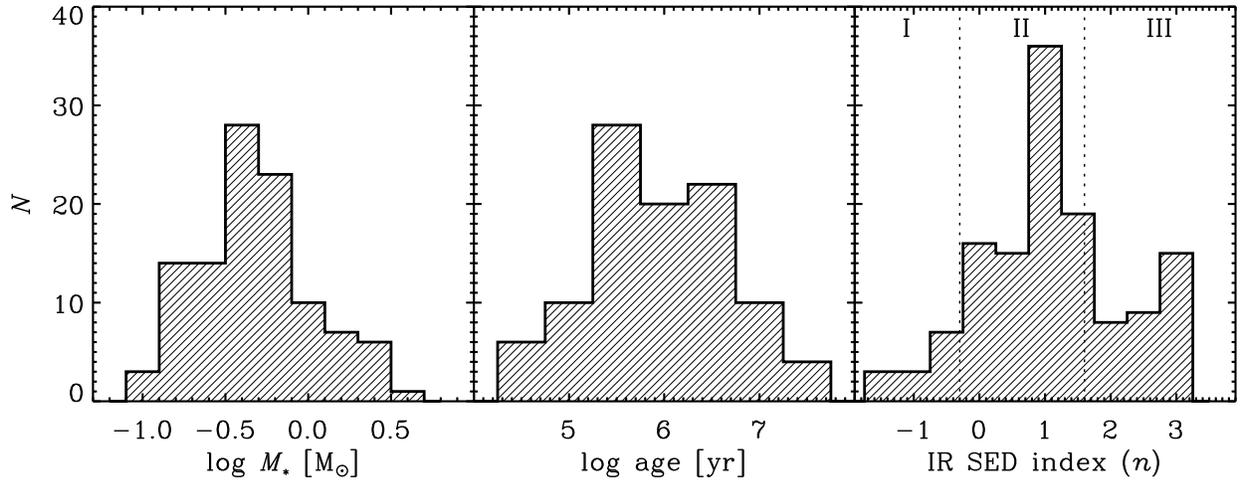}
\figcaption{Histograms of some relevant properties for the $\rho$ Oph sample 
objects.  The left and middle panels show stellar masses and ages, inferred as 
described in \S 4.2 using the \citet{dantona97} theoretical pre-main$-$sequence 
evolution models.  The right panel shows the infrared SED index (defined as $n$ 
where $\nu F_{\nu} \propto \nu^n$; see \S 3.3) that is commonly used as an 
evolutionary diagnostic for circumstellar dust.  YSOs with an index less than 
$-0.3$ are Class I sources, between $-0.3$ and 1.6 are Class II sources, and 
$>$1.6 are Class III sources \citep[SED classification borders are shown as 
dotted vertical lines;][]{greene94}.  \label{sampleprops}}
\end{figure}

Unfortunately, the large submillimeter photometry survey of $\rho$ Oph YSOs 
that was originally planned was cut short by the de-commissioning of SCUBA in 
the summer of 2006.  For a more complete sample of $\rho$ Oph similar in size 
and content to that presented in Paper I for Taurus, the data described above 
were supplemented with a large number of 1.3\,mm single-dish observations from 
the literature \citep{am94,jensen96,nurnberger98,motte98,stanke06}.  This 
expanded sample consists of 147 objects; 48 with multiwavelength data, and 99 
others with only 1.3\,mm flux densities from previous work.  The vast majority 
of the sources lie in the L1688 dark cloud, with a few others scattered in 
L1689 and the surrounding Oph-Sco complex.  Table \ref{results_table} lists the 
350, 450, 850, and 1300\,$\mu$m flux densities for these 147 sources, along 
with disk masses (\S 3.1), submillimeter colors (\S 3.2), and other relevant 
properties.  Figure \ref{sampleprops} summarizes some of the basic 
characteristics of the $\rho$ Oph sample.  

In what follows, we follow \citet{wilking05} and adopt a distance of 150\,pc to 
$\rho$ Oph.  For this distance, the projected FWHM beam sizes are 1350\,AU for 
both 350 and 450\,$\mu$m and 2100\,AU for 850\,$\mu$m.  Because these 
single-dish beams are substantially larger than the circumstellar dust 
associated with Class II and III sources \citep[e.g.,][]{kitamura02,andrews07}, 
it is safe to assume that the values listed in Table \ref{results_table} for 
such objects represent the total integrated continuum flux densities.  As shown 
in mapping observations \citep[e.g.,][]{motte98,stanke06} and discussed in 
Paper I, this is not likely the case for Class I sources.  It is important to 
emphasize that the photometry for the Class I YSOs listed in Table 
\ref{results_table} corresponds to the combined continuum emission 
contributions from both a circumstellar disk and the inner regions of a 
circumstellar envelope.

\section{Results}

\subsection{Disk Masses}

Following our work in Paper I, we aim to provide some empirical calibration of 
the relationship between submillimeter luminosities and disk masses through the 
use of a simple disk structure model.  A geometrically thin irradiated dust 
disk emits thermal radiation modified by the optical properties of the solid 
particles from a continuous series of annuli at a range of stellocentric 
radii.  The SED of this emission at a given frequency $\nu$ (here defined as 
$L_{\nu} = 4 \pi d^2 \nu F_{\nu}$) can be described in terms of the physical 
conditions in the disk as
\begin{equation}
L_{\nu} = 8 \pi^2 \nu \cos{i} \int_{r_0}^{R_d} B_{\nu}(T_r) \left( 1 - e^{-\tau_{\nu,r}\sec{i}} \right) \, r \,\, dr,
\end{equation}
where $i$ is the disk inclination, $r_0$ and $R_d$ the inner and outer radii, 
$B_{\nu}(T_r)$ the Planck function at a radius-dependent temperature, and 
$\tau_{\nu,r}$ the optical depth of the disk material 
\citep[e.g.,][]{adams87,beckwith90,chiang97}.  The latter is taken to be the 
product of the opacity, $\kappa_{\nu}$, and surface density, $\Sigma_r$.  The 
disk structure is approximated with radial power-laws, where $T_r \propto 
r^{-q}$ and $\Sigma_r \propto r^{-p}$.  To simplify the SED modeling process, 
we define the opacity to have a spatially homogeneous power-law spectrum such 
that $\kappa_{\nu} \propto \nu^{\beta}$ with $\beta = 1$ and a normalization of 
$\kappa_0 = 0.1$\,cm$^2$ g$^{-1}$ at 1000\,GHz \citep[cf.,][]{beckwith90}.  
This opacity implicitly assumes a 100:1 mass ratio of gas to dust.

We identified a subsample of 19 $\rho$ Oph YSOs that had suitable mid-infrared 
through millimeter SED data in the literature to determine $M_d$ from a fit to 
this simple disk structure model.  For consistency with the work done in Paper 
I for Taurus sources, we fixed the following parameters in the SED modeling of 
this subsample: $i = 0$\degr, $r_0 = 0.01$\,AU, $R_d = 100$\,AU, and $p = 
1.5$.  The SEDs longward of 5\,$\mu$m were de-reddened (see \S 4.2) and fit to 
the model using $\chi^2$-minimization.  The results of the SED fits are 
compiled in Table \ref{sedfits_table}, including literature sources for the SED 
data, the reduced $\chi^2$ statistic, degrees of freedom in the fit, and the 
best-fit values for the 3 free model parameters; $T_1$ (the temperature at 
1\,AU), $q$ (the radial power-law index of the temperature profile), and $M_d$ 
(determined from the surface density profile and fixed outer radius).

\begin{figure}
\epsscale{1.0}
\plotone{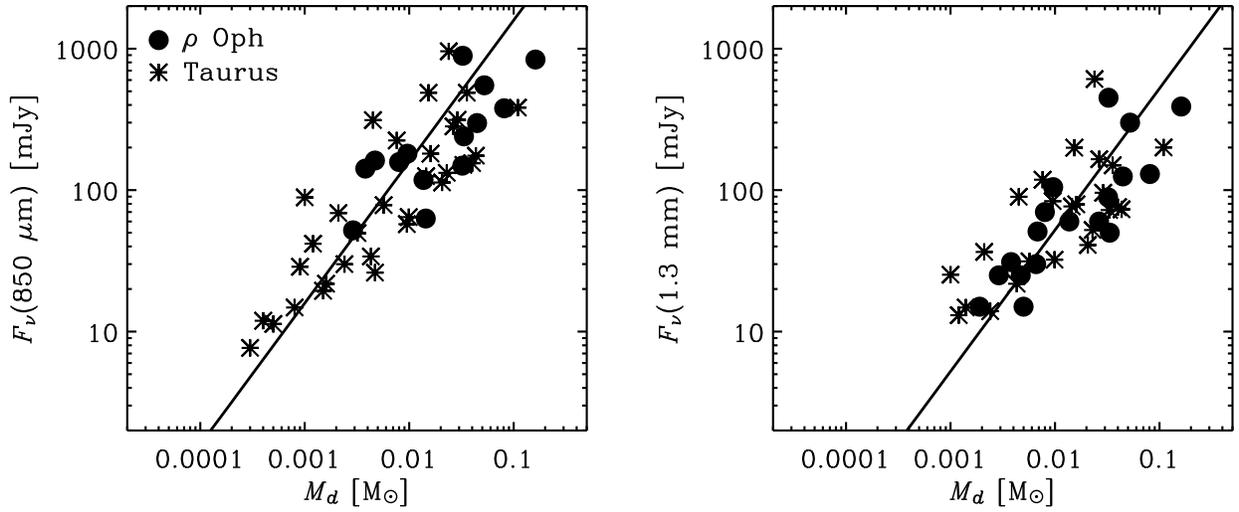}
\figcaption{Correlations between the 850\,$\mu$m ({\it left}) or 1.3\,mm ({\it 
right}) flux densities and the best-fit disk masses inferred from fitting SEDs 
with a simple disk structure model.  The results for $\rho$ Oph ({\it circles}: 
see Table \ref{sedfits_table}) and Taurus ({\it asterisks}: see Paper I) YSOs 
are shown together (the latter flux densities have been scaled to $d=150$\,pc 
for ease of comparison).  Overlaid as a solid line in each panel is the generic 
relationship described by Eqn.~2 for $T_c = 20$\,K, used to calculate disk 
masses (or upper limits) for the remainder of the sample.  \label{mf_cals}}
\end{figure}

Using these results and their counterparts for the Taurus sample (see Paper I), 
Figure \ref{mf_cals} demonstrates that the submillimeter flux densities and 
best-fit disk masses are correlated.  Most of the YSOs do not have sufficient 
data to determine $M_d$ from a SED fit, and therefore an alternative method is 
required.  As discussed in the introduction, most disks are primarily optically 
thin at submillimeter wavelengths \citep{beckwith90}.  Assuming that 
$\tau_{\nu,r} \ll 1$ and the emission is generated in a roughly isothermal 
region with a characteristic temperature ($T_c$), Eqn.~1 reduces to
\begin{equation}
L_{\nu} \approx 8 \pi^2 \nu \kappa_{\nu} B_{\nu}(T_c) M_d.
\end{equation}
This approximation explicitly shows that the submillimeter luminosity is a 
direct diagnostic of the total optical depth in the disk, or the product 
$\kappa_{\nu} M_d$.  The behavior of this simplified relationship is able to 
reproduce the SED fit results for both Taurus and $\rho$ Oph YSOs quite well in 
Figure 2, where it is plotted as a solid line for $T_c = 20$\,K.  

\begin{figure}
\epsscale{0.5}
\plotone{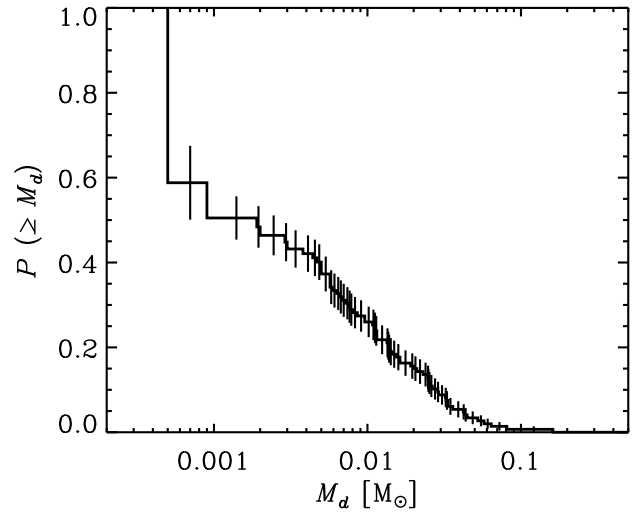}
\figcaption{The cumulative distribution of disk masses for the $\rho$ Oph 
sources listed in Table \ref{results_table}, constructed using the Kaplan-Meier 
estimator to include non-detections.  The ordinate represents the probability 
that a sample disk has a mass greater than the abscissa value.  The median disk 
mass for the full sample is $\sim$0.002\,M$_{\odot}$. \label{Md_CDF}}
\end{figure}

The disk masses (or upper limits) for the remaining 128 YSOs in this sample 
without sufficient SED data were computed according to Eqn.~2 with a 
characteristic temperature of 20\,K.  These $M_d$ values are compiled in Table 
\ref{results_table}, with preference given to using the more modern and 
sensitive 850\,$\mu$m flux densities when available.  The resulting cumulative 
distribution of $\rho$ Oph disk masses for the full sample is shown in Figure 
\ref{Md_CDF}, constructed with the Kaplan-Meier product-limit estimator to 
incorporate the upper limits \citep[\citealt{feigelson85}; see also e.g., Paper 
I or][]{am94}.  This distribution (and others like it in what follows) shows 
the probability that a given disk in the sample has a mass greater than or 
equal to the abscissa value.  The median disk mass in the full $\rho$ Oph 
sample is $\langle M_d \rangle \approx 0.002$\,M$_{\odot}$.  Roughly 28\%\ of 
these sources have $M_d \ge 0.01$\,M$_{\odot}$, a representative value for the 
minimal amount of solar-composition material required to account for the 
planets in the Solar System \citep[the Minimum Mass Solar Nebula, hereafter 
MMSN;][]{weidenschilling77}.  Only 40\%\ of the $\rho$ Oph YSOs have $M_d 
\lesssim 0.001$\,M$_{\odot}$, corresponding to the mass of Jupiter.

\subsection{Submillimeter Colors}

The submillimeter continuum spectra of YSOs generally follow a simple power-law 
behavior, such that $F_{\nu} \propto \nu^{\alpha}$.  The spectral index 
$\alpha$ is essentially a submillimeter color, and potentially encapsulates 
important information about the optical properties of solid particles in the 
disk.  For the optically thin, isothermal approximation (Eqn.~2), the 
submillimeter continuum emission has a spectral dependence $L_{\nu} \propto 
\nu^3 \kappa_{\nu} \propto \nu^{3+\beta}$ in the Rayleigh-Jeans limit 
\citep{beckwith90,beckwith91}.  With these assumptions, the submillimeter color 
is directly related to the shape of the opacity spectrum, $\alpha = 2 + \beta$, 
which in turn is thought to be a composite diagnostic of the dust particle size 
distribution, mineralogical composition, and morphology 
\citep[e.g.,][]{miyake93,henning96,draine06}.  In the ISM, the generally small 
(sub-micron) dust particles are relatively inefficient submillimeter emitters 
with $\beta \approx 1.7$ \citep[e.g.,][]{li01}, resulting in steep 
submillimeter spectra with $\alpha \approx 4$ \citep[e.g.,][]{hildebrand83}.  
However, in the high-density environment of a circumstellar disk, solid 
particles are expected to grow to substantially larger sizes via collisional 
agglomeration \citep[see the recent review by][]{dominik07}.  For a given grain 
composition and morphology, increased particle sizes due to such growth will 
decrease the value of $\beta$ \citep[e.g.,][]{miyake93,draine06}.  This in turn 
would be manifested as a shallow submillimeter continuum spectrum, or decreased 
$\alpha$.

\begin{figure}
\epsscale{0.5}
\plotone{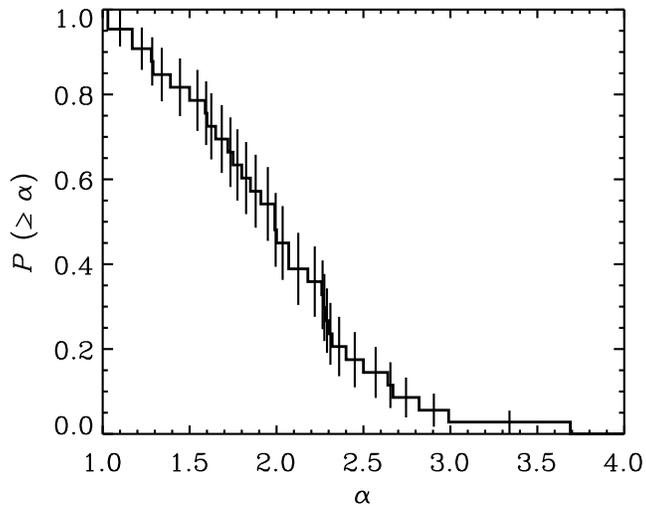}
\figcaption{The cumulative distribution of submillimeter colors ($\alpha$, 
defined as $F_{\nu} \propto \nu^{\alpha}$) for the $\rho$ Oph sources listed in 
Table \ref{results_table}, constructed using the Kaplan-Meier estimator to 
include non-detections.  The ordinate represents the probability that a sample 
disk has a spectral index greater than the abscissa value.  The median value 
for the full sample is $\langle \alpha \rangle \approx 2.0$. \label{n_CDF}}
\end{figure}

To assess the potential signatures of grain growth in $\rho$ Oph disks, we have 
calculated submillimeter colors ($\alpha$) between 350\,$\mu$m and 1.3\,mm 
whenever multiwavelength data were available.  In cases with $>$2 flux density 
measurements in this wavelength range, $\alpha$ was determined by a linear fit 
in the $\{\log{\nu}, \log{F_{\nu}}\}$ plane.  Otherwise, a simple flux ratio 
was utilized to measure a color or upper limit.  The inferred values of 
$\alpha$ are included in Table \ref{results_table}.  Figure \ref{n_CDF} shows 
their cumulative distribution, again constructed with the Kaplan-Meier 
estimator to include upper limits.  The median value is $\langle \alpha \rangle 
\approx 2$, with only a few percent of the YSOs showing $\alpha \ge 3$.  

Based on the above discussion, it would be tempting to conclude that the 
submillimeter SEDs for the $\rho$ Oph sample show clear evidence for 
significant solid particle growth (with $\beta \lesssim 1$).  However, there 
are a number of other effects that can serve to mimic the signatures of a 
shallow opacity spectrum.  Perhaps the most relevant of these is optically 
thick contamination, originating in the dense inner disk.  As a tracer of the 
disk temperature, emission with large optical depths will bias $\alpha$ to 
lower values because it has the same spectral shape as the Planck function in 
the Rayleigh-Jeans limit ($\alpha = 2$).  Interferometric observations 
that resolve the disk structure can be used to quantify the contamination, and 
usually indicate optically thick fractions of $\lesssim20-30$\%\ shortward of 
1\,mm \citep[e.g.,][]{testi01,testi03,andrews07}.  Moreover, the presumed 
origin of the submillimeter continuum emission near the cold disk midplane 
\citep[e.g.,][]{chiang97,dullemond02} implies that the Rayleigh-Jeans limit may 
not be valid (i.e., if much of the emission comes from dust with $T \lesssim 
15$\,K).  In this case, the thermal emission is considerably shallower than 
$\nu^2$ and may contribute to the observed low $\alpha$ values.  Finally, there 
are more technical considerations, particularly the rather large uncertainties 
on $\alpha$ due to calibration inaccuracies and small frequency leverage, as 
well as contamination from non-disk emission (see Paper I for details).  

Despite these uncertainties, it would be difficult to account for the 
submillimeter color distribution in Figure \ref{n_CDF} for an opacity spectrum 
similar to that in the ISM \citep[see][]{dalessio01}.  Detailed disk structure 
modeling and spatially resolved data are required to reach a more definitive 
conclusion.  In those cases where the disk structure has been resolved and can 
be used to correct the colors ($\alpha$) for optically thick contamination, the 
typical opacity index is $\beta \approx 1$ 
\citep{testi01,testi03,rodmann06,lommen07,andrews07}.  Without resolved data, 
the colors presented here serve solely as empirical properties of YSOs.  While 
the opacity spectra of these sources remain uncertain, the observed colors are 
certainly significantly different than noted for dust in the ISM, in the same 
sense as expected for when particle sizes are increased.  The issue of grain 
growth will be addressed further in \S 4.3.

\subsection{A Comparison with Disks in Taurus-Auriga}

Having established the basic submillimeter properties of YSOs in $\rho$ Oph, we 
can make a comparison with their counterparts in Taurus from Paper I.  Taurus 
is a substantially less crowded region, with projected stellar densities 
reaching only $\sim$50$-$60 stars pc$^{-2}$ \citep{gomez93} compared with the 
$\sim$200 stars pc$^{-2}$ noted in the $\rho$ Oph core \citep{greene92}.  
Despite the higher concentration of YSOs in $\rho$ Oph, a simple calculation of 
the collision timescale shows that dynamical interactions are not expected to 
significantly affect their disk properties.  De-projecting the \citet{greene92} 
stellar density into a 1\,pc$^{3}$ volume ($n_{\ast} \sim 200$\,pc$^{-3}$) and 
assuming a typical disk size ($r \sim 100$\,AU $\Rightarrow \sigma \sim 
10^{-8}$\,pc$^{-2}$, where $\sigma$ is the collisional cross section) and 
cluster velocity dispersion ($v \sim 1$\,km s$^{-1}$) implies a disk$-$disk 
collision timescale of $t_c \approx (n_{\ast} \sigma v)^{-1} \sim 1$\,Gyr, 
roughly 3 orders of magnitude larger than the cluster lifetime.  

In addition to being more spatially concentrated, $\rho$ Oph members are often 
still deeply embedded in their natal molecular cloud material.  This is in 
stark contrast to most YSOs in the Taurus association, which by comparison are 
relatively unobscured.  Given these environmental distinctions, it is 
worthwhile to examine if they have any impact on disk properties by directly 
comparing the results for the two regions.  

\begin{figure}
\epsscale{0.5}
\plotone{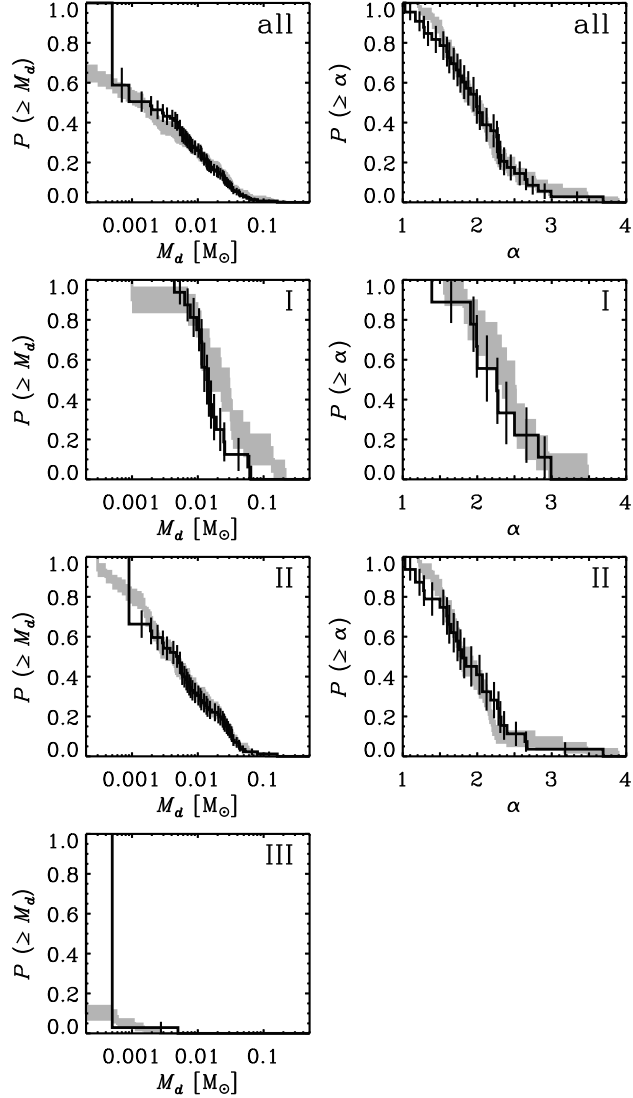}
\figcaption{Direct comparisons of submillimeter properties in the $\rho$ Oph 
(black lines) and Taurus (grayscale) regions.  The panels on the left and right 
show the cumulative distributions of $M_d$ and $\alpha$, respectively, for 
various subgroups of YSOs in each region.  The top panels show the full samples 
described here and in Paper I.  The subsequent panels below show the 
distributions for Class I, II, and III sources.  The distributions of these 
submillimeter properties in the $\rho$ Oph and Taurus regions are essentially 
indistinguishable.  However, there are statistically significant changes in the 
$M_d$ and $\alpha$ distributions along the Class I $\rightarrow$ II 
$\rightarrow$ III evolution sequence.  \label{compare_CDFs}}
\end{figure}

To make the comparison a fair one, most of the Taurus disk masses were 
re-calculated from the 850\,$\mu$m flux densities in Paper I using Eqn.~2, the 
same parameters adopted here (see \S 3.1), and the appropriate distance, $d = 
140$\,pc.  A subsample of Taurus sources have sufficient information to 
determine $M_d$ from a fit to the full SED (see Table 2 in Paper I), and in 
those cases the best-fit disk masses were adopted.  Direct comparisons of the 
cumulative distribution functions of $M_d$ and $\alpha$ for the similar-sized 
($\sim$150 YSOs) $\rho$ Oph and Taurus samples are shown in the top panels of 
Figure \ref{compare_CDFs}.  These distribution functions take into account 
upper limits using the Kaplan-Meier estimator and should be a reliable 
comparison for $M_d \gtrsim 0.001$\,M$_{\odot}$, corresponding to a rough 
luminosity limit for the $\rho$ Oph sample.  The error envelopes of the Taurus 
distributions are shown in grayscale and the $\rho$ Oph distributions as 
histograms with error bars.  The two-sample statistical tests described in the 
survival analysis formalism of \citet[][i.e., the Gehan, Logrank, Peto \& Peto, 
and Peto \& Prentice tests]{feigelson85} confirm what is apparent from Figure 
\ref{compare_CDFs}: the distributions of $M_d$ and $\alpha$ for $\rho$ Oph and 
Taurus YSOs are essentially indistinguishable.  The results of these tests are 
provided in the top half of Table \ref{stats_compare}.

Since the relative numbers of YSOs in various evolutionary states are not the 
same in the $\rho$ Oph and Taurus samples, a closer look at more similar 
subsamples makes a better comparison.  The shape of the infrared SED is often 
utilized as an empirical diagnostic of the YSO evolution sequence 
\citep{lada84,adams86,adams87}.  The parameter of interest is the SED index 
$n$, defined from $\sim$2$-$25\,$\mu$m via $\nu F_{\nu} \propto \nu^n$.  We 
adopt the classification breakdown suggested by \citet{greene94}, where Class I 
objects (star+disk+envelope) have $n \le -0.3$, Flat-Spectrum objects 
(star+disk+envelope?) have $-0.3 < n < 0.3$, Class II objects (star+disk) have 
$0.3 \le n \le 1.6$, and Class III objects (star only) have $n > 1.6$.  
Infrared photometry for the $\rho$ Oph sample was collected from the literature 
\citep{wilking89,weaver92,greene94,bontemps01,barsony05,padgett06}, the 2MASS 
Point Source Catalog \citep{cutri03}, and the preliminary results of the 
``Cores to Disks" {\it Spitzer} Legacy Survey of the region (L.~Allen 2007, 
private communication).  The infrared photometry of Taurus sources utilized for 
this same purpose in Paper I was supplemented with {\it Spitzer} photometry 
from \citet{hartmann05} and \citet{luhman06}.  Infrared SED slopes ($n$) were 
computed for each YSO in the $\rho$ Oph and Taurus samples with a simple 
power-law fit to the SED, corrected for reddening whenever possible 
(extinctions were computed as described in \S 4.2 or Paper I).  The resulting 
SED classifications for the $\rho$ Oph sample are included in Table 
\ref{results_table}.

This empirical evolution sequence was used to generate YSO subsamples in both 
star-forming environments and compare them with the aforementioned two-sample 
statistical tests.  The results are compiled in the top portion of Table 
\ref{stats_compare}, and the cumulative distributions of $M_d$ and $\alpha$ for 
various SED classifications are also shown in Figure \ref{compare_CDFs}.  The 
$P$ values in Table \ref{stats_compare} correspond to the probability that the 
two subsamples are drawn from {\it different} parent distributions.  The ranges 
of $P$ represent the various statistical tests \citep[see][]{feigelson85}.  
After concluding that the few Flat-Spectrum sources do not substantially 
influence the results (as was noted in Paper I), they were incorporated into 
the Class II subsample.  

Although not significant by formal statistical arguments, the distributions of 
$M_d$ and $\alpha$ for Class I objects in the two regions appear somewhat 
different in Figure \ref{compare_CDFs}.  Little import should be attached to 
this apparent discrepancy, as the photometry has not been conducted in a 
systematic way and the potential for different amounts of envelope 
contamination has therefore not been properly taken into account.  On the other 
hand, the level of detail in the agreement between the $\rho$ Oph and Taurus 
distributions of $M_d$ and $\alpha$ for Class II sources is remarkable.  For 
disk masses greater than $\sim$0.001\,M$_{\odot}$, the distributions between 
the two regions are essentially identical, even showing the same dip feature 
around the MMSN value (0.01\,M$_{\odot}$) and decrease to virtually zero 
probability for disks with $M_d \gtrsim 0.1$\,M$_{\odot}$.  The median Class II 
disk in both samples has $\langle M_d \rangle \approx 0.005$\,M$_{\odot}$, with 
roughly one third of each sample harboring disks with $M_d \ge M_{\rm MMSN}$.  
The shapes of the submillimeter color distributions between the two regions are 
also fairly precise matches, with median values $\langle \alpha \rangle \approx 
2$ and only a few percent of each sample having $\alpha \ge 3$.  

As with the full samples, the subsamples described above indicate that there 
are no statistically significant differences between the submillimeter 
properties of YSOs in the $\rho$ Oph and Taurus regions at {\it any given 
evolutionary state}.  Given the roughly equivalent sizes and basic 
characteristics of the samples in these two regions and the homogeneous 
methodology utilized to infer $M_d$ and $\alpha$, this result clearly indicates 
that the local environment of a low-mass, low stellar density ($\ll 
10^3$\,stars pc$^{-2}$) region does not substantially affect the submillimeter 
properties of the associated disks.

\section{Discussion}

\subsection{Disk Evolution}

While confirming the similarities between the submillimeter properties of 
circumstellar dust in the $\rho$ Oph and Taurus regions, the panels in Figure 
\ref{compare_CDFs} also show that there are substantial changes in $M_d$ and 
$\alpha$ across the empirical evolution sequence characterized by the infrared 
SED index.  Two-sample tests on the censored datasets directly confirm that the 
$M_d$ distributions of Class I, II, and III sources in $\rho$ Oph are 
statistically different.  The test results are catalogued in the bottom half of 
Table \ref{stats_compare}.  As with the YSOs in the Taurus clouds (Paper I), 
submillimeter luminosities (and therefore disk masses) and colors decrease 
substantially along the Class I $\rightarrow$ II $\rightarrow$ III sequence.  

\begin{figure}
\epsscale{0.5}
\plotone{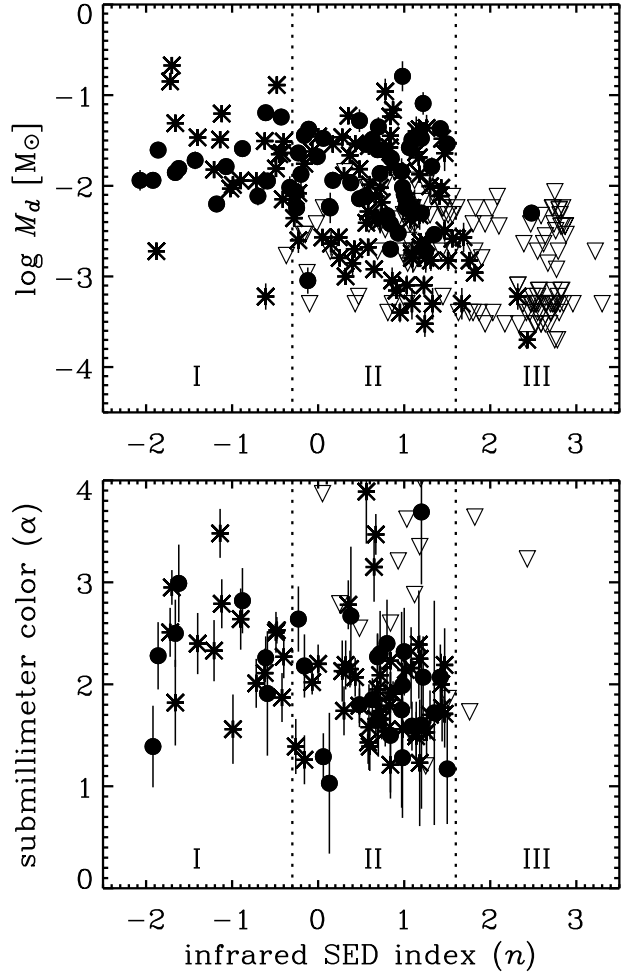}
\figcaption{Individual disk masses ({\it top}) and submillimeter colors 
($\alpha$: {\it bottom}) as a function of the infrared SED index measured 
between $\sim$2$-$25\,$\mu$m ($n$) for the combined $\rho$ Oph ({\it circles}) 
and Taurus ({\it asterisks}) samples.  Open triangles represent upper limits.  
Both panels show correlations such that $M_d$ and $\alpha$ decrease along the 
empirical evolution sequence (as $n$ increases).  \label{alpha_smm}}
\end{figure}

Figure \ref{alpha_smm} directly demonstrates how the disk masses and 
submillimeter colors vary as a function of the infrared SED index ($n$), now 
combining the data for the $\rho$ Oph and Taurus samples.  The top panel 
reinforces what is seen in the left panels of Figure \ref{compare_CDFs}: $M_d$ 
decreases as the infrared SED becomes bluer (i.e., $n$ increases) along the 
evolution sequence.  Using only the detections, a correlation between $M_d$ and 
$n$ is present at the 4\,$\sigma$ level, with a Spearman rank correlation 
coefficient of $-0.32$.  The correlation significance is enhanced when the 
upper limits are incorporated with any of the survival analysis statistics 
discussed by \citet{isobe86}.  Disk masses decrease smoothly with a shallow 
slope for YSOs with infrared SEDs redward of the Class II/III boundary (at $n 
\sim 1.6$), but then drop precipitously.  This non-linear behavior in the $\{n, 
M_d\}$-plane was in fact originally postulated in the pioneering study by 
\citet{am94}, along with the first indications that the $\rho$ Oph and Taurus 
disk mass distributions were the same.  In an effort to approximately determine 
the depth of the drop-off beyond the Class III boundary, we stacked the 
850\,$\mu$m SCUBA photometry observations of all undetected Class III sources 
in $\rho$ Oph and Taurus (Paper I) to derive their average luminosity value 
\citep[see, e.g.,][]{carpenter02}.  The mean 850\,$\mu$m flux density for the 
54 stacked Class III non-detections (5 in $\rho$ Oph and 49 in Taurus) was 
$0.1\pm0.3$\,mJy.  With the same assumptions adopted in \S 3.1, the 3\,$\sigma$ 
``average" limit (0.9\,mJy) indicates that the typical undetected Class III 
source has $M_d \lesssim 5\times10^{-5}$\,M$_{\odot}$, roughly the mass of 
Neptune.  

The bottom panel in Figure \ref{alpha_smm} examines the comparative evolution 
of colors in the infrared and submillimeter bands.  There is a marginal 
(2.9\,$\sigma$) correlation between $\alpha$ and $n$ in this diagram with a 
Spearman rank correlation coefficient of $-0.30$, suggesting that $\alpha$ 
systematically decreases along the infrared evolution sequence.  A similar 
correlation for the Taurus sources alone was noted in Paper I, and there 
partially attributed to a genuinely evolving disk opacity.  The idea behind 
this argument is that disks are expected to become more optically thin as YSOs 
evolve along the infrared SED sequence.  In the case of no opacity evolution, a 
decreasing fraction of optically thick submillimeter emission as $n$ increases 
should lead to increasing $\alpha$, the opposite of what is noted in Figure 
\ref{alpha_smm}.  On the other hand, the growth of solid particles as the 
circumstellar dust evolves along the sequence would result in a flattening of 
the opacity spectrum (decreased $\beta$), and therefore the observed decreasing 
$\alpha$.  An alternative (or possibly cooperative) scenario could be a 
decreasing temperature in the disk midplane \citep[e.g., due to gravitational 
sedimentation of dust to the disk midplane;][]{dalessio06}.  These 
possibilities can potentially be explored with spatially resolved 
multiwavelength data, although with such a weak correlation many more sources 
would need to be observed with interferometers.  

The above analysis confirms for $\rho$ Oph what was shown for Taurus in Paper 
I: the infrared and submillimeter properties of circumstellar dust around young 
stars evolve along a similar empirical sequence.  Given where such emission 
originates, this result suggests that the material in the inner (infrared) and 
outer (submillimeter) parts of disks is significantly affected by global 
changes in disk properties.  One of the obvious candidates for the cause of 
such evolution is the metamorphosis of circumstellar structure, including the 
dissipation of the envelope (Class I $\rightarrow$ II), the viscous evolution 
of the accretion disk (through Class II), and the dissipation of the disk 
(Class II $\rightarrow$ III).  A perhaps equally important contribution to this 
evolution could be the result of the substantial growth of solid particles.  In 
reality, the submillimeter luminosity is determined by the product 
$\kappa_{\nu} M_d$: even if the total circumstellar mass remained the same 
across the evolution sequence, the observed trends could both be explained by 
the changes in the opacity expected from grain growth.  

Given the abruptness of the transition at the Class II/III boundary in the top 
panel of Figure \ref{alpha_smm} and the low submillimeter detection rate for 
Class III sources (see also Paper I), it appears that the signatures of the 
inner and outer disk (infrared and submillimeter excesses, respectively) 
disappear within a short time of one another.  In Paper I, we used the standard 
``duty-cycle" argument for this based on the very small percentage 
($\lesssim 3$\%) of YSOs that have long-wavelength (i.e., submillimeter) 
emission from the outer disk, but no near-infrared excess from the inner disk.  
The scarcity of such ``transition" objects suggests that if all disks go 
through such a phase, it must be rather short in duration ($\le 10^5$\,yr).  A 
variety of mechanisms have been proposed for this rapid radial evolution, from 
planet formation \citep[see][and references therein]{najita07} to rapid mass 
loss in photoevaporative winds \citep{clarke01,alexander06}.

\subsection{Relation to Stellar Properties}

Despite the above effort to associate changes in the submillimeter properties 
of circumstellar dust with an empirical evolution sequence, it is still not 
possible to tie this sequence to an absolute timeframe.  An alternative 
examination of disk evolution can be made by relating disk and stellar 
parameters, particularly masses ($M_{\ast}$) and ages ($t_{\ast}$).  To 
determine the latter, we first collected spectral classifications (see Table 
\ref{results_table}) and optical/near-infrared photometry from the literature 
\citep{rydgren80,chini81,myers87,herbig88,bouvier92,vrba93,herbst94,greene95,luhman99,wilking99,chavarria00,prato03,doppmann03,doppmann05,eisner05,wilking05,gras-velazquez05,padgett06,grankin07}.  For sources in the surveys of 
\citet{wilking05} or \citet{doppmann03,doppmann05}, we simply adopted their 
stellar masses and ages.  For the remainder of the sample, we used the spectral 
type$-$effective temperature scale, intrinsic colors, and bolometric 
corrections tabulated by \citet{kh95}.  Visual extinctions were calculated from 
color excesses (in the optical-red whenever available) and the \citet{cohen81} 
extinction law.  Stellar luminosities were then computed from the de-reddened 
photometry and appropriate bolometric correction.  Stellar masses and ages were 
determined by reference to the \citet{dantona97} theoretical 
pre$-$main-sequence mass tracks and isochrones in an H-R diagram.  

\begin{figure}
\epsscale{0.5}
\plotone{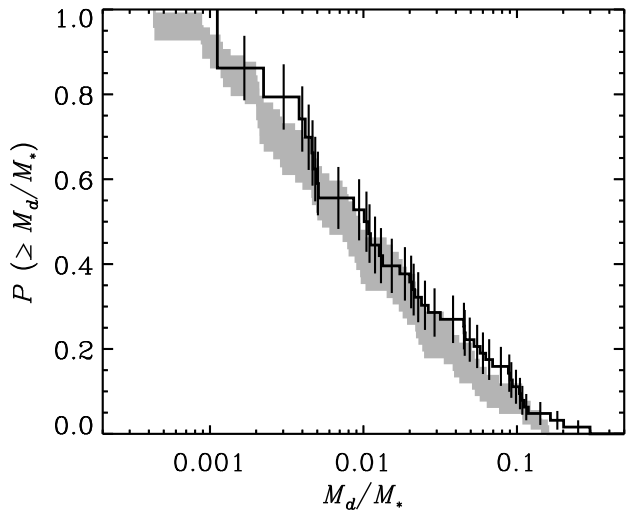}
\figcaption{The cumulative distributions of the disk-to-star mass ratio for 
Class II sources in $\rho$ Oph ({\it lines}) and Taurus ({\it grayscale}).  
While the $\rho$ Oph sources appear to generally have slightly larger mass 
ratios than their Taurus counterparts, two-sample tests on the censored data 
indicate that the difference is not statistically significant.  The median 
ratios are $\sim$1.0\%\ and 0.8\%\ for $\rho$ Oph and Taurus, respectively. 
\label{CDF_MdMs}}
\end{figure}

This is the same technique adopted for the Taurus YSOs in Paper I, and is 
similar in practice to that utilized by \citet{wilking05}.  The stellar mass 
and age distributions were already highlighted in Figure \ref{sampleprops}.  
The $\rho$ Oph sample is slightly younger than the Taurus sample, with a median 
age of $\sim$0.7\,Myr compared to $\sim$1\,Myr.  The cumulative distributions 
of disk-to-star mass ratios ($M_d/M_{\ast}$) for Class II sources in both the 
$\rho$ Oph (lines) and Taurus (grayscale) samples are shown together in Figure 
\ref{CDF_MdMs}.  The $\rho$ Oph sources generally exhibit slightly higher mass 
ratios than their Taurus counterparts, although the $M_d/M_{\ast}$ 
distributions for the two regions are not formally different in a statistical 
sense.  The median ratio in $\rho$ Oph is $\langle M_d/M_{\ast} \rangle \approx 
1.0$\%, compared to $\sim$0.8\%\ for Taurus.  In this case, selection effects 
may be mimicking a subtle diagnostic of disk evolution from the slightly 
younger $\rho$ Oph sample to the more mature Taurus YSOs.  The $\rho$ Oph 
sample does have a somewhat different $M_{\ast}$ distribution, suggesting that 
any true hint of a change in $M_d/M_{\ast}$ as a function of cluster age should 
await larger samples with more similarity in their stellar properties.

\begin{figure}
\epsscale{0.45}
\plotone{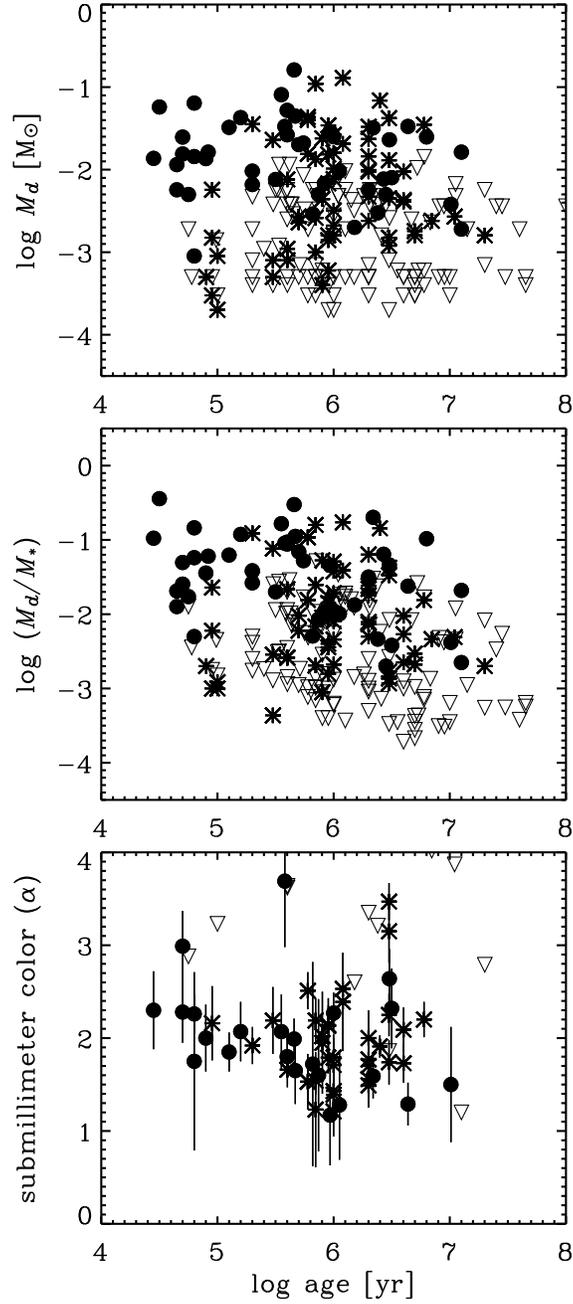}
\figcaption{Disk masses, disk-to-star mass ratios, and submillimeter colors as 
a function of the stellar age for the combined $\rho$ Oph ({\it circles}) and 
Taurus ({\it asterisks}) samples.  Open triangles are upper limits.  
Correlation tests on the censored data show that the disk-to-star mass ratio 
decreases significantly with time, although similar trends for $M_d$ and 
$\alpha$ are not clear.  \label{age_smm}}
\end{figure}

Figure \ref{age_smm} exhibits the variations of $M_d$, $M_d/M_{\ast}$, and 
$\alpha$ with time for the combined $\rho$ Oph and Taurus sample.  While there 
is a large scatter of disk properties at any given time, some marginal trends 
are apparent.  Although a definitive correlation between $M_d$ and $t_{\ast}$ 
remains unconfirmed, there are clearly more non-detections as the stars get 
older.  Using survival analysis techniques to incorporate upper limits, the Cox 
proportional hazard model, generalized Kendall's tau, and Spearman's rho 
correlation tests \citep{isobe86} all suggest that the disk-to-star mass ratio 
is indeed correlated with the stellar age, dropping from a few percent at 
$\sim$0.1\,Myr to a factor of 30 or more lower by 10\,Myr.  It would be useful 
to measure disk masses for clusters with slightly {\it older} ages to confirm 
such trends, so that the strength of the correlation does not hinge on the 
youngest sources (i.e., $t \lesssim 10^5$\,yr), where the isochrones are less 
trustworthy.  No obvious trend between $\alpha$ and age is apparent.  

Before moving on, it is worthwhile to comment on the adopted distance ($d = 
150$\,pc) and how alternative values would be expected to affect the above 
analysis.  Following \citet{wilking05} and others, we based the choice of $d$ 
on the typical range of $\sim$130$-$170\,pc from measurements of stars in the 
larger Upper Scorpius complex that encompasses the $\rho$ Oph clouds 
\citep[e.g.,][]{whittet74,degeus89,knude98,dezeeuw99,mamajek07}.  However, 
other authors have focused on the near end of that range, using the $d = 
125$\,pc advocated by de Geus and colleagues based on two arguments suggesting 
that the dark clouds are in the Upper Sco foreground: ($a$) the morphology of 
the distance modulus vs. extinction diagram for Upper Sco sources near the 
$\rho$ Oph cloud edge \citep{degeus89}, and ($b$) some circumstantial evidence 
from the morphology of the local ISM \citep{degeus92}.  

If this closer distance is adopted, all luminosities utilized above are reduced 
by $\sim$30\%, with two important consequences.  First, the decreased 
bolometric luminosities in the H-R diagram would result in considerably older 
ages for the $\rho$ Oph sample (stellar masses would remain approximately the 
same, owing to the essentially vertical tracks in this diagram).  With the 
roughly logarithmic behavior of isochrones along the H-R diagram luminosity 
axis, the increase in $t_{\ast}$ could be a factor of 2$-$4.  Second, this 
nearer distance would lead to a substantial discrepancy in the $\rho$ Oph and 
Taurus $M_d$ distributions, such that disks in the former region would be 
significantly less massive.  Taken together, using a distance as close as 
125\,pc would imply that $\rho$ Oph YSOs are {\it older} than their Taurus 
counterparts, and that disk masses have decreased substantially over the 
corresponding age difference.  Obviously using ages from the H-R diagram is not 
an ideal way to constrain a cluster distance.  However, in this case it does 
not really make sense that the much more deeply embedded YSOs in $\rho$ Oph are 
older than their essentially unobscured Taurus counterparts.  Perhaps this 
issue can be resolved with a more complete census of the stellar population in 
the $\rho$ Oph clouds, along the lines of the work done by \citet{wilking05} 
and \citet{doppmann03,doppmann05}.

\subsection{Benchmark Submillimeter Properties of Circumstellar Disks}

\begin{figure}
\epsscale{0.6}
\plotone{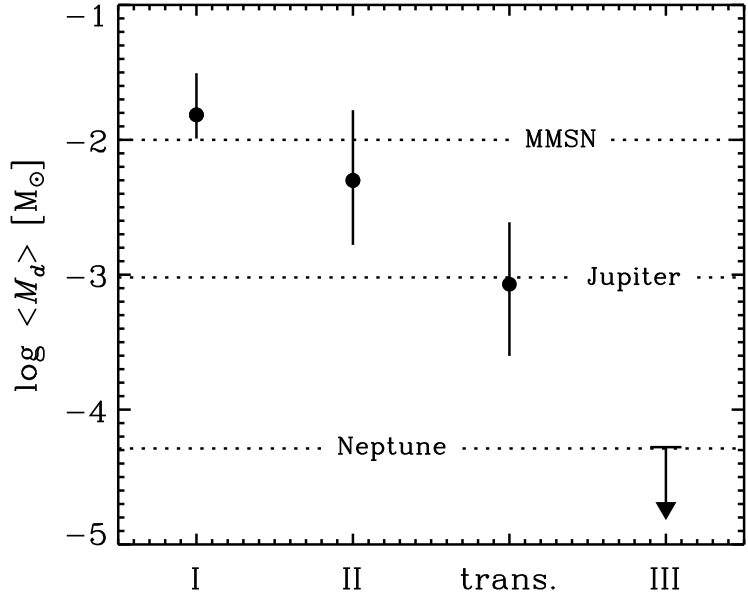}
\figcaption{Schematic diagram showing the evolution of the median disk masses 
across the empirical sequence defined by the shape of the infrared SED (see 
also Table \ref{benchmark_table}).  The transition objects (``trans.") are 
defined as Class III sources with firm submillimeter detections, and the Class 
III limit corresponds to the stacking analysis value determined in \S 4.1.  
Dashed horizontal lines mark standard reference masses.  \label{evolution}}
\end{figure}

Now that large submillimeter photometry samples of YSOs for the $\rho$ Oph and 
Taurus associations have been collected, it is appropriate to present 
quantitative reference values for some of the basic properties of their 
circumstellar environments.  The cumulative distributions of the censored 
datasets presented above were used to determine the median values of $M_d$, 
$M_d/M_{\ast}$, and $\alpha$ compiled in Table \ref{benchmark_table} for 
various evolutionary states.  Here we have distinguished the Class III sources 
($n > 1.6$) into two categories: those with submillimeter detections are termed 
``transition" objects (because dust in the inner disk is largely absent, while 
a significant outer disk remains) and those undetected sources retain the 
standard Class III label.  These benchmark values clearly highlight the 
evolution in the submillimeter properties that have been carefully determined 
in this work and Paper I.  Figure \ref{evolution} graphically illustrates the 
trend of decreasing $\langle M_d \rangle$ along this sequence, with 
representative error bars showing the distribution quartiles.  Typical disk 
masses are similar to the MMSN for much of their evolution, only dropping 
substantially at the end of the Class II stage.  The transition disks have a 
typical disk mass roughly an order of magnitude lower than the MMSN value, 
corresponding roughly to a Jupiter mass of material.  The stringent stacked 
upper limit placed on the remaining Class III disks is more than an order of 
magnitude lower still, suggesting that primordial circumstellar disks at this 
stage in evolution are probably non-existent.  However, this limit still does 
not rule out the presence of low-mass debris disks similar to those noted 
around older nearby stars \citep[e.g.,][]{holland98}.

\begin{figure}
\epsscale{1.00}
\plotone{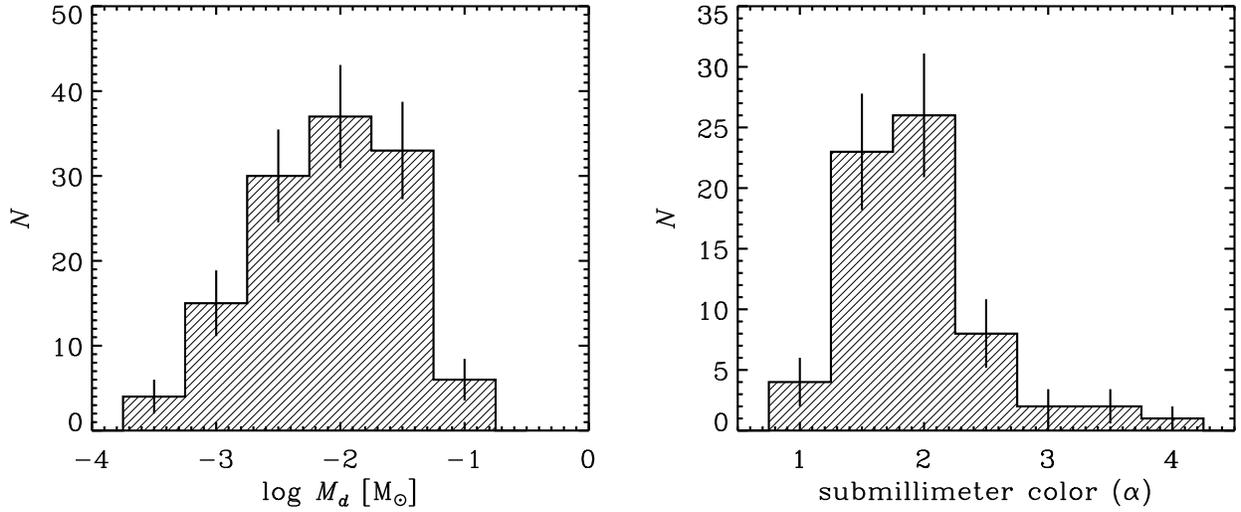}
\figcaption{The differential distributions of $M_d$ and $\alpha$ for Class II 
sources that were firmly detected at submillimeter wavelengths in the combined 
$\rho$ Oph and Taurus surveys.  Uncertainties correspond to the counting errors 
in each bin.  These distributions are representative of the known properties of 
disks that may represent the initial conditions for the planet formation 
process.  \label{benchmarks}}
\end{figure}

The accretion disks present in the Class II stage of evolution are generally 
regarded as the birthsites of planetary systems.  As such, their properties are 
directly relevant for constraining the initial conditions available for the 
planet formation process.  In light of their significance in this matter, 
Figure \ref{benchmarks} shows the differential $M_d$ and $\alpha$ distributions 
for the $\rho$ Oph and Taurus Class II sources that have firm submillimeter 
detections.  Compiled from 125 individual sources, this distribution of disk 
masses peaks at $\sim$0.01\,M$_{\odot}$ (the MMSN value) and covers a fairly 
wide dispersion.  As would be expected from the analysis of individual regions 
presented here and in Paper I, the $\alpha$ distribution for these sources has 
a relatively narrow peak centered around $\alpha \approx 2$.  With the 
similarities between the disk properties in the two star-forming regions (\S 
3.3), these values should represent a combined, universal description of the 
most basic submillimeter properties of Class II disks.  

But how precisely are the values of these properties constrained in an absolute 
sense?  Of course, the submillimeter colors are empirically measured quantities 
and, aside from occasionally large error bars, their values are firmly 
determined.  The challenge is interpreting the $\alpha$ values in terms of the 
solid particle population in these disks.  The evidence is now fairly clear on 
the observational side that the typical disk opacity spectrum is more shallow 
than in the ISM, with $\beta \sim 1$ deemed appropriate to explain the 
submillimeter SEDs \citep[e.g., Paper 
I;][]{beckwith91,mannings94,rodmann06,andrews07}.  This effect has been 
interpreted as the signature of an increase in the maximum dust particle size, 
from $\sim$sub-micron sizes in the ISM to $\sim$millimeter sizes in a typical 
disk \citep{miyake93,pollack94,henning96,dalessio01,draine06}.  This amounts 
to $\sim$4 orders of magnitude in particle growth from the time when these dust 
grains were incorporated into circumstellar disks/envelopes from the ISM.  
Moreover, \citet{dalessio01} emphasize that this actually represents a lower 
limit on the amount of particle growth, as $\beta$ remains near unity for 
maximum particle sizes well beyond the (sub-)millimeter wavelengths used to 
constrain it (see their Fig.~3).  In essence, submillimeter colors indicate 
that significant grain growth has occurred in these disks, although it remains 
unclear exactly how far along it may have proceeded beyond millimeter sizes.

In a sense, the opacity acts as an accounting system for the mass budget in a 
disk.  Not all solid particle sizes are generating sufficiently detectable 
emission at submillimeter wavelengths, and the opacity must also represent the 
mass contribution of the inefficient emitters in the conversion of a 
submillimeter luminosity to a disk mass.  A direct, non-degenerate measurement 
of $\kappa_{\nu}$ in a disk is not currently possible, and so a standard 
approximate value has been fixed and regularly utilized in this field.  In 
their admittedly simplified analysis, \citet{dalessio01} show that this fixed 
opacity value, $\kappa_{{\rm 1.3 mm}} \approx 0.02$\,cm$^{2}$ g$^{-1}$ (as well 
as $\beta = 1$), is appropriate when the largest particle sizes are 
$\sim$1\,mm.  However, the opacity is significantly diminished when the maximum 
particle size is larger than this value.  Because $M_d \propto 
L_{{\rm smm}}/\kappa_{\nu}$, an underestimate of the amount of particle growth 
in these disks implies an underestimate of their masses.

\begin{figure}
\epsscale{0.5}
\plotone{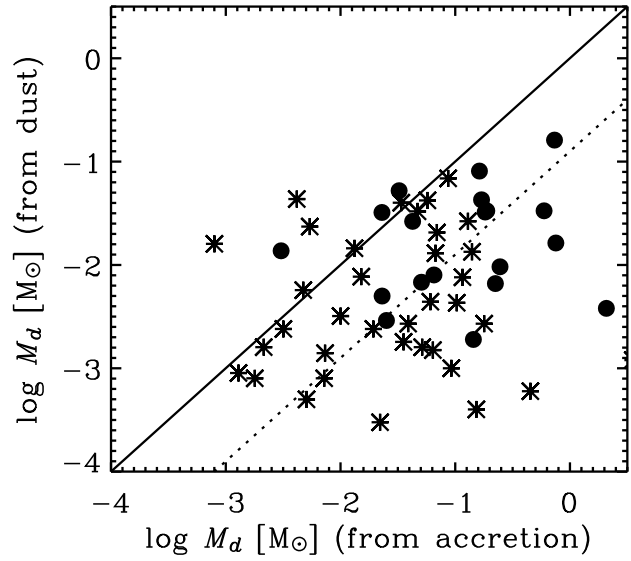}
\figcaption{A comparison of disk masses inferred in two independent ways for 
$\rho$ Oph ({\it circles}) and Taurus ({\it asterisks}) Class II disks.  The 
abscissa uses gas accretion rates from optical/infrared spectra and some 
assumptions about the accretion history \citep[$M_d \approx 2 
t_{\ast}$\emph{\.M};][]{hartmann98,natta06} and the ordinate the dust masses 
described in \S 3.1 and Paper I.  The solid line marks the expected behavior if 
the values were equal.  The dotted line bisects the sample, and indicates that 
dust masses underestimate the accretion masses by nearly an order of 
magnitude.  Particle growth up to meter size scales may explain the 
discrepancy, due to a significant over-estimate of the disk opacity.  
\label{M_gasdust}} 
\end{figure}

Using laboratory experiments and computer models as a guide for the collision 
conditions in the disk interior, \citet{dominik07} argue that solid bodies up 
to $\sim$meter size scales can be formed on rather short timescales.  The work 
by \citet{dalessio01} shows that meter-sized bodies would have an opacity 
$\sim$10$\times$ lower than what has been assumed here (but with the same value 
of $\beta$).  If such growth is typical, the inferred disk masses may be 
underestimated by an order of magnitude.  \citet{hartmann98} proposed an 
interesting test on the validity of the standard opacity prescription.  Aside 
from the method used here based on the submillimeter continuum emission from 
dust, they argued that mass accretion rates (\emph{\.M}) determined from 
optical/ultraviolet spectra can be used to estimate $M_d$ with some reasonable 
assumptions about the accretion history.  Their disk mass estimates, $M_d 
\approx 2 t_{\ast}$\emph{\.M}, are based solely on observations of the gas 
phase in the disks, and are therefore completely independent of the values 
presented here \citep[see][regarding the numerical factor of 2]{hartmann98}.  
Using the stellar ages derived in the previous section and the accretion rates 
compiled by \citet{hartmann98} and \citet{natta06}, disk masses were calculated 
using both methods for the subset of $\rho$ Oph and Taurus Class II sources 
with the required information available (i.e., \emph{\.M}, $t_{\ast}$, and a 
millimeter detection).\footnote{We preferred to re-calculate the accretion 
rates provided by \citet{natta06} based on the stellar parameters adopted 
throughout this paper.  Regardless, the results in Figure \ref{M_gasdust} stand 
irrespective of whose numbers are used.}  

A comparison of the two methods for computing $M_d$ in Figure \ref{M_gasdust} 
clearly demonstrates that disk masses estimated from dust emission are 
systematically smaller than those inferred from the accretion rates.  The 
dotted line in this plot has a slope of unity and bisects the sample into 
equal-sized groups: it lies roughly an order of magnitude below the solid line 
marking equality between the two $M_d$ determination methods.  Of course, both 
methods of estimating $M_d$ involve significant uncertainties.  The unknown 
disk structure in most cases can lead to relatively small uncertainties on the 
$M_d$ values determined from submillimeter continuum emission (see Paper I; 
Andrews \& Williams 2007).  Larger uncertainties plague the $M_d$ estimates 
from the \citet{hartmann98} method, due to both the subtleties of determining 
{\emph{\.M}} and the challenges of specifying individual stellar ages.  While 
these undoubtedly contribute to the scatter in Figure \ref{M_gasdust}, they are 
unlikely to cause the noted {\it systematic} offset.  For example, a 
temperature uncertainty could easily translate to a 50\%\ $M_d$ uncertainty 
(see Eqn.~2) for a given disk.  But, the scatter in Figure \ref{mf_cals} shows 
that it is improbable that the average adopted $T_c$ is systematically 
2$\times$ lower than assumed (let alone the 10$\times$ needed to explain Figure 
\ref{M_gasdust}).  A similar argument holds for other disk structure parameters 
and the stellar ages used in determining $M_d$ from the accretion rate.  

Perhaps a variety of uncertainties can unfortunately conspire to produce the 
observed offset.  To be fair, this comparison of $M_d$ estimates from different 
material phases and spatial locations (gas near the inner rim and dust spread 
to the outer radius) relies on the assumption that accretion disks can be 
approximated with relatively simple physical prescriptions.  However, given all 
of these caveats, Figure \ref{M_gasdust} is really the only check on the 
absolute precision of $M_d$ measurements currently available.  If solely 
physical explanations are considered to explain these results, the growth of 
solid particles to centimeter or meter size scales (rather than the millimeter 
scales that correspond to the standard opacity prescription) is perhaps the 
most likely cause.  Such growth is observationally reinforced by the remarkable 
centimeter-wave dust emission from large particles recently detected for the TW 
Hya disk \citep{wilner05}.  

As discussed above, the decrease in the opacity expected from this level of 
grain growth corresponds to disk masses being under-estimated by up to an order 
of magnitude.  If this is the case, the implications for the planet formation 
process are significant.  The appropriate shift in the $M_d$ distributions 
presented here to account for the $\kappa_{\nu}$ over-estimate could imply that 
a relatively large fraction (perhaps up to one third) of $\sim$1\,Myr old Class 
II disks are massive enough to be gravitationally unstable \citep[i.e., 
$M_d/M_{\ast} \gtrsim 0.2$;][]{shu90}.  If these instabilities are indeed 
common, the giant planet formation process may be rapidly accelerated via 
direct condensations from spiral density waves/rings 
\citep[e.g.,][]{boss98,mayer07}, compared to the standard core accretion and 
gas capture models \citep[e.g.,][]{pollack96,hubickyj05}.  A concerted effort 
utilizing both high spatial and spectral resolution multiwavelength 
submillimeter data (e.g., from the ALMA and e-VLA interferometers) and 
state-of-the-art disk models may help shed some light on this subject in the 
near future.

\section{Summary}

We have presented new multiwavelength submillimeter continuum observations of 
48 YSOs in the $\rho$ Oph dark clouds.  Supplementing this survey with 99 
millimeter measurements in the literature, we study the circumstellar dust 
around young stars in this region; the results are summarized here.
\begin{itemize}
\item The statistical distributions of disk masses ($M_d$), disk-to-star mass 
ratios, and submillimeter colors ($\alpha$; where $F_{\nu} \propto 
\nu^{\alpha}$) were derived and directly compared with those for the Taurus 
association \citep{aw05}.  Within the uncertainties, there are no significant 
differences between the submillimeter properties of YSOs in the two regions.
\item There are statistically significant decreases in $M_d$ and $\alpha$ along 
the canonical YSO evolution sequence based on the shape of the infrared SED.  
Changes in the structure of the circumstellar environment and/or the average 
particle properties may explain the correlations.
\item A tentative correlation between the disk-to-star mass ratio and stellar 
age are noted for the combined samples in both regions.  Confirmation with 
older sources would be beneficial to assess whether or not this relationship 
depends on isochrone models at early ages ($\sim 10^5$\,yr).
\item The Class II YSOs in $\rho$ Oph and Taurus are utilized to establish 
benchmark values for submillimeter properties in the population considered to 
be the birthsites of planetary systems: $M_d \sim 0.005$\,M$_{\odot}$, 
$M_d/M_{\ast} \sim 1$\%, and $\alpha \sim 2$.  
\item An independent disk mass estimate based on the accretion rate 
\citep{hartmann98} is used to compare with the submillimeter continuum 
measurements.  The latter are systematically smaller than the former by roughly 
an order of magnitude on average.  This may be the result of substantial 
particle growth (up to $\sim$meter sizes), leading to an over-estimate of the 
opacity used to compute $M_d$ from the submillimeter luminosity.  Such disk 
mass underestimates may imply that some fraction of Class II disks are 
marginally gravitationally unstable, which could have important implications 
for disk evolution and giant planet formation.
\end{itemize}

\acknowledgments
We acknowledge useful conversations and advice from Lee Hartmann, Eric Mamajek, 
and an anonymous referee.  We would especially like to thank Lori Allen, Rob 
Gutermuth, and their colleagues for providing very useful {\it Spitzer} data 
before publication, as well as the JCMT and CSO support staffs for their 
assistance.  This work was supported by the NASA Graduate Student Researchers 
Program (GSRP) grant NNG05G012H and the NSF grant AST-0324328.  This research 
has made use of the NASA/IPAC Infrared Science Archive, which is operated by 
the Jet Propulsion Laboratory, California Institute for Technology, under 
contract with the National Aeronautics and Space Administration.

\clearpage

\begin{deluxetable}{lcccccccrc}
\tablecolumns{10} 
\tabletypesize{\scriptsize}
\tablewidth{0pc}
\tablecaption{Submillimeter Properties of Circumstellar Disks in $\rho$ Ophiuchus \label{results_table}}
\tablehead{
\colhead{} & \colhead{} & \colhead{} & \colhead{$F_{\nu}$(350\,$\mu$m)} & \colhead{$F_{\nu}$(450\,$\mu$m)} & \colhead{$F_{\nu}$(850\,$\mu$m)} & \colhead{$F_{\nu}$(1.3\,mm)} & \colhead{$M_d$} & \colhead{} &\colhead{} \\ \colhead{Object} & \colhead{SED} & \colhead{SpT} & \colhead{[mJy]} & \colhead{[mJy]} & \colhead{[mJy]} & \colhead{[mJy]} & \colhead{[M$_{\odot}$]} & \colhead{$\alpha$} & \colhead{Notes} \\ \colhead{(1)} & \colhead{(2)} & \colhead{(3)} & \colhead{(4)} & \colhead{(5)} & \colhead{(6)} & \colhead{(7)} & \colhead{(8)} & \colhead{(9)} & \colhead{(10)}} 
\startdata
Wa Oph 1, HBC 630   & III & K2      & \nodata      & $< 114$      & $< 8$      & $< 20$     & $< 0.0005$ & \nodata       & 1 \\
AS 205, HBC 254     & II  & K5      & $4683\pm118$ & $3280\pm271$ & $891\pm11$ & $450\pm10$ & 0.03       & $1.85\pm0.21$ & 1 \\
Wa Oph 2, HBC 633   & III & K1      & \nodata      & $<129$       & $< 9$      & $< 20$     & $< 0.0006$ & \nodata       & 1 \\
Wa Oph 3, HBC 634   & III & K0      & \nodata      & $<145$       & $< 10$     & $< 25$     & $< 0.0006$ & \nodata       & 1 \\
WSB 3               & III & M3      & \nodata      & \nodata      & $< 34$     & \nodata    & $< 0.002$  & \nodata       & 2 \\
WSB 4               & III & M3      & \nodata      & \nodata      & \nodata    & $< 13$     & $< 0.003$  & \nodata       & 3 \\
WSB 11              & II  & \nodata & \nodata      & \nodata      & $< 26$     & \nodata    & $< 0.001$  & \nodata       & 2 \\
L1719 B, 16191-1936 & FS  & \nodata & \nodata      & \nodata      & \nodata    & $220\pm10$ & 0.04       & \nodata       & 1 \\
WSB 18              & III & M2      & \nodata      & \nodata      & $< 46$     & \nodata    & $< 0.003$  & \nodata       & 2 \\
WSB 19              & II  & M3      & \nodata      & \nodata      & $< 78$     & \nodata    & $< 0.004$  & \nodata       & 2 \\
DoAr 16, HBC 257    & II  & K6      & \nodata      & $< 361$      & $47\pm8$   & $< 50$     & 0.003      & $< 3.21$      & 1 \\
SR 1, EL 9, GSS 9   & III & B2      & \nodata      & $< 154$      & $< 11$     & $< 50$     & $< 0.0007$ & \nodata       & 1 \\
SR 22, DoAr 19      & II  & M4      & \nodata      & $< 162$      & $31\pm3$   & $< 20$     & 0.002      & $< 2.60$      & 1 \\
ROXs 2, VSS 20      & III & M0      & \nodata      & $< 156$      & $< 11$     & $< 30$     & $< 0.0007$ & \nodata       & 1 \\
IRS 2, YLW 19       & II  & K3      & \nodata      & \nodata      & \nodata    & $< 25$     & $< 0.005$  & \nodata       & 1 \\
IRS 7               & II  & \nodata & \nodata      & \nodata      & \nodata    & $< 40$     & $< 0.008$  & \nodata       & 1 \\
IRS 8               & III & \nodata & \nodata      & \nodata      & \nodata    & $< 40$     & $< 0.008$  & \nodata       & 1 \\
IRS 9, YLW 23       & III & \nodata & \nodata      & \nodata      & \nodata    & $< 10$     & $< 0.002$  & \nodata       & 1 \\
ROXs 3, HBC 636     & III & M0      & \nodata      & \nodata      & \nodata    & $< 30$     & $< 0.006$  & \nodata       & 1 \\
IRS 10, ROXs 4      & III & K6      & \nodata      & \nodata      & \nodata    & $< 28$     & $< 0.005$  & \nodata       & 3 \\
ROXs 5              & III & K7      & \nodata      & \nodata      & \nodata    & $< 19$     & $< 0.004$  & \nodata       & 3 \\
SR 4, AS 206        & II  & K5      & \nodata      & $< 768$      & $142\pm7$  & $31\pm6$   & 0.004      & $1.50\pm0.62$ & 3 \\
IRS 13, ROXs 7      & III & K6      & \nodata      & \nodata      & \nodata    & $< 20$     & $< 0.004$  & \nodata       & 1 \\
DoAr 21, HBC 637    & III & K1      & \nodata      & \nodata      & \nodata    & $< 30$     & $< 0.006$  & \nodata       & 4 \\
SR 3, IRS 16        & III & A0      & \nodata      & \nodata      & \nodata    & $< 25$     & $< 0.005$  & \nodata       & 1 \\
GSS 26              & II  & K8      & \nodata      & $650\pm123$  & $298\pm7$  & $125\pm20$ & 0.04       & $1.65\pm0.36$ & 4 \\
EL 18, GSS 29       & II  & K6      & \nodata      & \nodata      & \nodata    & $< 10$     & $< 0.002$  & \nodata       & 4 \\
DoAr 24, HBC 638    & II  & K5      & \nodata      & \nodata      & \nodata    & $< 30$     & $< 0.006$  & \nodata       & 4 \\
CRBR 12, ISO-Oph 21 & I   & \nodata & \nodata      & \nodata      & \nodata    & $85\pm5$   & 0.02       & \nodata       & 4 \\
DoAr 23, WSB 26     & III & M0      & \nodata      & \nodata      & $< 77$     & \nodata    & $< 0.004$  & \nodata       & 2 \\
EL 20, VSSG 1       & II  & \nodata & \nodata      & $< 666$      & $240\pm7$  & $50\pm10$  & 0.03       & $3.69\pm0.71$ & 1 \\
CRBR 15             & I   & M5      & \nodata      & \nodata      & \nodata    & $40\pm5$   & 0.008      & \nodata       & 4 \\
EL 21, GSS 30       & I   & \nodata & \nodata      & $1403\pm116$ & $222\pm5$  & $90\pm30$  & 0.01       & $2.50\pm0.33$ & 4 \\
LFAM 1              & FS  & \nodata & \nodata      & $2662\pm219$ & $572\pm8$  & $250\pm20$ & 0.04       & $2.18\pm0.31$ & 4 \\
GY 10               & III & M9      & \nodata      & \nodata      & \nodata    & $< 10$     & $< 0.002$  & \nodata       & 4 \\
GY 11               & FS  & M7      & \nodata      & \nodata      & \nodata    & $< 10$     & $< 0.002$  & \nodata       & 4 \\
DoAr 24E, GSS 31    & II  & K1      & \nodata      & $779\pm121$  & $158\pm6$  & $70\pm20$  & 0.008      & $2.32\pm0.43$ & 4 \\
DoAr 25, WSB 29     & II  & K5      & \nodata      & \nodata      & $461\pm11$ & $280\pm10$ & 0.03       & $1.17\pm0.54$ & 1 \\
LFAM 3, GY 21       & FS  & \nodata & \nodata      & \nodata      & \nodata    & $110\pm10$ & 0.02       & \nodata       & 4 \\
ROXs 9A             & III & M0      & \nodata      & \nodata      & \nodata    & $< 22$     & $< 0.004$  & \nodata       & 3 \\
EL 24, WSB 31       & II  & K6      & $5242\pm213$ & $3012\pm216$ & $838\pm8$  & $390\pm10$ & 0.2        & $1.99\pm0.22$ & 1 \\
ROXs 12             & III & M0      & \nodata      & \nodata      & \nodata    & $< 19$     & $< 0.004$  & \nodata       & 3 \\
VSSG 27, GY 51      & II  & \nodata & \nodata      & \nodata      & \nodata    & $< 20$     & $< 0.004$  & \nodata       & 4 \\
ROXs 9B             & III & K2      & \nodata      & \nodata      & \nodata    & $< 11$     & $< 0.002$  & \nodata       & 3 \\
IRS 14, GY 54       & II  & \nodata & \nodata      & \nodata      & \nodata    & $< 30$     & $< 0.006$  & \nodata       & 1 \\
IRS 15, GY 58       & II  & \nodata & \nodata      & \nodata      & \nodata    & $< 30$     & $< 0.006$  & \nodata       & 1 \\
S1, EL 25, GSS 35   & III & B3      & \nodata      & \nodata      & \nodata    & $< 10$     & $< 0.002$  & \nodata       & 4 \\
ROXs 9C             & III & K4      & \nodata      & \nodata      & \nodata    & $< 14$     & $< 0.003$  & \nodata       & 3 \\
GY 91               & I   & \nodata & \nodata      & \nodata      & \nodata    & $300\pm15$ & 0.06       & \nodata       & 4 \\
WSB 37, GY 93       & II  & M5      & \nodata      & \nodata      & \nodata    & $< 25$     & $< 0.005$  & \nodata       & 1 \\
WL 8, GY 96         & II  & \nodata & \nodata      & \nodata      & \nodata    & $< 10$     & $< 0.002$  & \nodata       & 4 \\
IRS 17, WL 7        & FS  & \nodata & \nodata      & \nodata      & \nodata    & $< 30$     & $< 0.006$  & \nodata       & 4 \\
ROC 16, GY 101      & FS  & \nodata & \nodata      & \nodata      & \nodata    & $< 15$     & $< 0.003$  & \nodata       & 4 \\
EL 26, GSS 37       & II  & M0      & \nodata      & \nodata      & $< 51$     & $15\pm5$   & 0.005      & $< 2.88$      & 1 \\
IRS 19, VSSG 11     & III & M0      & \nodata      & \nodata      & \nodata    & $< 15$     & $< 0.003$  & \nodata       & 1 \\
WL 12, YLW 2        & I   & \nodata & \nodata      & $749\pm130$  & $216\pm7$  & $115\pm10$ & 0.01       & $2.00\pm0.36$ & 4 \\
EL 27, GSS 39       & II  & K8      & \nodata      & $2737\pm322$ & $678\pm10$ & $300\pm10$ & 0.04       & $2.07\pm0.32$ & 1 \\
ROXs 9D             & III & \nodata & \nodata      & \nodata      & \nodata    & $< 35$     & $< 0.007$  & \nodata       & 3 \\
WSB 38, VSS 27      & III & G4      & \nodata      & \nodata      & \nodata    & $< 20$     & $< 0.004$  & \nodata       & 1 \\
DoAr 28, HBC 261    & II  & K5      & \nodata      & \nodata      & \nodata    & $< 75$     & $< 0.01$   & \nodata       & 1 \\
WL 2, GY 128        & FS  & \nodata & \nodata      & \nodata      & \nodata    & $30\pm10$  & 0.006      & \nodata       & 4 \\
WL 18, GY 129       & II  & K7      & \nodata      & \nodata      & \nodata    & $85\pm10$  & 0.02       & \nodata       & 4 \\
IRS 23, WL 21       & III & \nodata & \nodata      & \nodata      & \nodata    & $< 30$     & $< 0.006$  & \nodata       & 4 \\
CRBR 51             & FS  & \nodata & \nodata      & \nodata      & \nodata    & $70\pm15$  & 0.01       & \nodata       & 4 \\
SR 24, HBC 262      & FS  & K1      & $1271\pm32$  & $1263\pm282$ & $530\pm9$  & $230\pm11$ & 0.03       & $1.29\pm0.23$ & 3 \\
WL 14, GY 172       & II  & M4      & \nodata      & \nodata      & \nodata    & $30\pm10$  & 0.006      & \nodata       & 4 \\
IRS 27, WL 22       & I   & \nodata & \nodata      & $< 438$      & $70\pm5$   & $40\pm10$  & 0.004      & $1.99\pm0.96$ & 4 \\
WL 16, YLW 5        & FS  & \nodata & \nodata      & \nodata      & \nodata    & $< 6$      & $< 0.001$  & \nodata       & 4 \\
IRS 29, WL 1        & FS  & \nodata & \nodata      & $< 702$      & $15\pm4$   & $< 20$     & 0.0009     & $< 6.05$      & 4 \\
LFAM 26, GY 197     & I   & \nodata & \nodata      & \nodata      & \nodata    & $100\pm10$ & 0.02       & \nodata       & 4 \\
WL 17, GY 205       & I   & \nodata & \nodata      & $< 867$      & $100\pm7$  & $75\pm10$  & 0.006      & $0.84\pm0.65$ & 4 \\
WL 10, GY 211       & II  & K8      & \nodata      & \nodata      & \nodata    & $< 60$     & $< 0.01$   & \nodata       & 4 \\
GY 213              & II  & \nodata & \nodata      & \nodata      & \nodata    & $40\pm10$  & 0.008      & \nodata       & 4 \\
EL 29, WL 15        & I   & \nodata & \nodata      & $2024\pm109$ & $408\pm6$  & $95\pm10$  & 0.03       & $2.82\pm0.32$ & 4 \\
SR 21, EL 30        & II  & G3      & $2761\pm57$  & $1896\pm268$ & $397\pm6$  & $95\pm15$  & 0.03       & $2.27\pm0.22$ & 4 \\
IRS 31, WL 9        & II  & \nodata & \nodata      & \nodata      & \nodata    & $< 10$     & $< 0.002$  & \nodata       & 4 \\
GY 224              & I   & \nodata & \nodata      & \nodata      & \nodata    & $50\pm10$  & 0.01       & \nodata       & 4 \\
WL 19, GY 227       & II  & \nodata & \nodata      & \nodata      & \nodata    & $30\pm10$  & 0.006      & \nodata       & 4 \\
WL 11, GY 229       & II  & \nodata & \nodata      & \nodata      & \nodata    & $< 10$     & $< 0.002$  & \nodata       & 4 \\
WSB 45, HBC 640     & III & M5      & \nodata      & \nodata      & \nodata    & $< 20$     & $< 0.004$  & \nodata       & 1 \\
IRS 33, GY 236      & II  & \nodata & \nodata      & \nodata      & \nodata    & $105\pm20$ & 0.02       & \nodata       & 4 \\
WSB 46, HBC 641     & II  & M2      & \nodata      & \nodata      & \nodata    & $< 20$     & $< 0.004$  & \nodata       & 1 \\
IRS 34, GY 239      & II  & \nodata & \nodata      & \nodata      & \nodata    & $< 20$     & $< 0.004$  & \nodata       & 4 \\
IRS 36, GY 241      & FS  & \nodata & \nodata      & \nodata      & \nodata    & $< 15$     & $< 0.003$  & \nodata       & 4 \\
WL 20, GY 240       & I   & \nodata & \nodata      & $400\pm112$  & $182\pm6$  & $47\pm1$   & 0.01       & $1.39\pm0.40$ & 5 \\
IRS 37, GY 244      & II  & M4      & \nodata      & $< 591$      & $93\pm8$   & $< 10$     & 0.006      & $1.03\pm0.69$ & 4 \\
IRS 38, WL 5        & III & F7      & \nodata      & $< 522$      & $80\pm4$   & $< 10$     & 0.005      & $< 4.89$      & 4 \\
GY 245              & FS  & \nodata & \nodata      & \nodata      & \nodata    & $60\pm5$   & 0.01       & \nodata       & 4 \\
IRS 39, WL 4        & II  & M2      & \nodata      & $< 480$      & $63\pm5$   & $< 15$     & 0.01       & $1.75\pm0.96$ & 4 \\
IRS 41, WL 3        & II  & \nodata & \nodata      & $< 582$      & $115\pm4$  & $< 60$     & 0.007      & $< 2.55$      & 4 \\
SR 12, HBC 263      & III & M0      & \nodata      & \nodata      & $< 26$     & $< 20$     & $< 0.001$  & \nodata       & 2,4 \\
IRS 42, GY 252      & II  & \nodata & \nodata      & \nodata      & \nodata    & $< 60$     & $< 0.01$   & \nodata       & 4 \\
WL 6, GY 254        & FS  & \nodata & \nodata      & \nodata      & \nodata    & $< 20$     & $< 0.004$  & \nodata       & 4 \\
VSSG 22, ROXs 23    & III & K6      & \nodata      & \nodata      & \nodata    & $< 30$     & $< 0.006$  & \nodata       & 1 \\
WSB 49              & II  & M4      & \nodata      & \nodata      & \nodata    & $< 25$     & $< 0.005$  & \nodata       & 1 \\
CRBR 85             & II  & \nodata & \nodata      & \nodata      & \nodata    & $150\pm10$ & 0.03       & \nodata       & 4 \\
YLW 16C, GY 262     & II  & M1      & \nodata      & \nodata      & \nodata    & $60\pm5$   & 0.03       & \nodata       & 4 \\
IRS 43, YLW 15      & I   & \nodata & \nodata      & $1811\pm358$ & $246\pm7$  & $80\pm10$  & 0.02       & $2.99\pm0.38$ & 4 \\
EL 31, WL 13        & II  & M0      & \nodata      & \nodata      & \nodata    & $< 10$     & $< 0.002$  & \nodata       & 4 \\
IRS 44, GY 269      & I   & \nodata & \nodata      & \nodata      & \nodata    & $60\pm10$  & 0.01       & \nodata       & 4 \\
EL 32, IRS 45       & II  & K7      & \nodata      & \nodata      & \nodata    & $< 50$     & $< 0.01$   & \nodata       & 4 \\
IRS 46, GY 274      & FS  & \nodata & \nodata      & \nodata      & \nodata    & $45\pm10$  & 0.009      & \nodata       & 4 \\
EL 33, IRS 47       & II  & \nodata & \nodata      & \nodata      & \nodata    & $< 20$     & $< 0.004$  & \nodata       & 4 \\
GY 284              & II  & M3      & \nodata      & \nodata      & \nodata    & $130\pm10$ & 0.02       & \nodata       & 4 \\
ROXs 25, GY 292     & II  & M2      & \nodata      & \nodata      & \nodata    & $30\pm5$   & 0.007      & \nodata       & 4 \\
ROXs 30A            & III & \nodata & \nodata      & \nodata      & \nodata    & $< 10$     & $< 0.002$  & \nodata       & 3 \\
IRS 48, GY 304      & I   & \nodata & \nodata      & $< 1566$     & $180\pm9$  & $60\pm10$  & 0.01       & $1.91\pm0.61$ & 4 \\
IRS 50, GY 306      & III & M4      & \nodata      & \nodata      & \nodata    & $< 10$     & $< 0.002$  & \nodata       & 4 \\
IRS 49, GY 308      & II  & K8      & \nodata      & $< 1041$     & $52\pm5$   & $25\pm5$   & 0.003      & $1.71\pm1.10$ & 1 \\
DoAr 32, WSB 51     & II  & K6      & \nodata      & \nodata      & \nodata    & $< 45$     & $< 0.009$  & \nodata       & 1 \\
DoAr 33, WSB 53     & II  & K4      & \nodata      & $< 678$      & $79\pm7$   & $40\pm10$  & 0.005      & $1.60\pm0.82$ & 1 \\
WSB 52, GY 314      & II  & K5      & \nodata      & \nodata      & \nodata    & $51\pm10$  & 0.007      & \nodata       & 6 \\
IRS 51, GY 315      & II  & \nodata & \nodata      & $< 678$      & $171\pm6$  & $110\pm10$ & 0.01       & $< 2.17$      & 4 \\
SR 9, AS 207        & II  & K5      & \nodata      & \nodata      & $< 25$     & $15\pm5$   & 0.002      & $< 1.20$      & 1 \\
EL 36, VSSG 14      & II  & A7      & \nodata      & \nodata      & \nodata    & $< 10$     & $< 0.002$  & \nodata       & 3 \\
IRS 54, GY 378      & FS  & \nodata & \nodata      & \nodata      & \nodata    & $30\pm10$  & 0.006      & \nodata       & 4 \\
ROXs 31, HBC 642    & III & K8      & \nodata      & \nodata      & $< 22$     & $< 25$     & $< 0.001$  & \nodata       & 1,2 \\
SR 10, HBC 265      & II  & M2      & \nodata      & \nodata      & \nodata    & $< 25$     & $< 0.005$  & \nodata       & 1 \\
WSB 60, YLW 58      & II  & M4      & $832\pm67$   & $651\pm161$  & $149\pm7$  & $89\pm2$   & 0.03       & $1.59\pm0.24$ & 5 \\
SR 20, HBC 643      & III & G7      & \nodata      & \nodata      & $< 25$     & $< 35$     & $< 0.002$  & \nodata       & 1 \\
SR 13, HBC 266      & II  & M4      & \nodata      & $740\pm200$  & $118\pm6$  & $60\pm10$  & 0.01       & $2.30\pm0.42$ & 1 \\
WSB 63              & II  & M2      & \nodata      & \nodata      & \nodata    & $< 25$     & $< 0.005$  & \nodata       & 1 \\
ROXs 35A            & III & K3      & \nodata      & \nodata      & \nodata    & $< 29$     & $< 0.006$  & \nodata       & 3 \\
VSS 44, ROXs 35B    & III & G5      & \nodata      & \nodata      & \nodata    & $< 19$     & $< 0.004$  & \nodata       & 3 \\
ROXs 39             & III & K6      & \nodata      & \nodata      & \nodata    & $< 30$     & $< 0.006$  & \nodata       & 1 \\
DoAr 42, HBC 267    & III & M0      & \nodata      & \nodata      & \nodata    & $< 30$     & $< 0.006$  & \nodata       & 1 \\
ROXs 42B            & III & M0      & \nodata      & \nodata      & \nodata    & $< 45$     & $< 0.009$  & \nodata       & 1 \\
ROXs 42C            & II  & K6      & \nodata      & \nodata      & \nodata    & $< 30$     & $< 0.006$  & \nodata       & 1 \\
ROXs 43A            & II  & G0      & \nodata      & \nodata      & \nodata    & $< 35$     & $< 0.007$  & \nodata       & 1 \\
ROXs 43B            & III & K5      & \nodata      & \nodata      & \nodata    & $< 17$     & $< 0.003$  & \nodata       & 3 \\
IRS 60              & II  & \nodata & \nodata      & \nodata      & \nodata    & $< 50$     & $< 0.01$   & \nodata       & 1 \\
DoAr 44, HBC 268    & II  & K3      & \nodata      & $< 414$      & $181\pm6$  & $105\pm11$ & 0.01       & $1.28\pm0.59$ & 3 \\
IRS 63, L1709 B     & I   & \nodata & $6703\pm162$ & $4729\pm199$ & $1017\pm7$ & $370\pm10$ & 0.06       & $2.28\pm0.33$ & 1 \\
ROXs 45B            & III & \nodata & \nodata      & \nodata      & \nodata    & $< 19$     & $< 0.004$  & \nodata       & 3 \\
ROXs 45C            & III & K5      & \nodata      & \nodata      & \nodata    & $< 19$     & $< 0.004$  & \nodata       & 3 \\
ROXs 45D            & III & K0      & \nodata      & \nodata      & \nodata    & $< 19$     & $< 0.004$  & \nodata       & 3 \\
IRS 67, L1689 S     & I   & \nodata & \nodata      & $1687\pm207$ & $394\pm9$  & $150\pm10$ & 0.02       & $2.28\pm0.33$ & 1 \\
DoAr 51, HBC 647    & III & M0      & \nodata      & \nodata      & \nodata    & $< 20$     & $< 0.004$  & \nodata       & 1 \\
IRS 69              & II  & \nodata & \nodata      & \nodata      & \nodata    & $< 75$     & $< 0.01$   & \nodata       & 1 \\
DoAr 52, HBC 648    & II  & M2      & \nodata      & \nodata      & \nodata    & $< 55$     & $< 0.01$   & \nodata       & 1 \\
RNO 90, HBC 649     & II  & G5      & \nodata      & $557\pm83$   & $162\pm4$  & $25\pm5$   & 0.005      & $2.40\pm0.43$ & 1 \\
RNO 91, HBC 650     & FS  & M1      & \nodata      & $1524\pm87$  & $365\pm9$  & $90\pm10$  & 0.02       & $2.64\pm0.32$ & 1 \\
EL 49               & II  & \nodata & \nodata      & \nodata      & \nodata    & $58\pm5$   & 0.01       & \nodata       & 3 \\
Wa Oph 4, HBC 652   & II  & K4      & \nodata      & \nodata      & \nodata    & $< 13$     & $< 0.003$  & \nodata       & 1 \\
Wa Oph 6, HBC 653   & II  & K6      & \nodata      & $979\pm297$  & $379\pm7$  & $130\pm10$ & 0.08       & $2.07\pm0.40$ & 1 \\
Wa Oph 5, HBC 654   & II  & M2      & \nodata      & \nodata      & \nodata    & $< 25$     & $< 0.005$  & \nodata       & 1 \\
AS 209, HBC 270     & II  & K5      & $3000\pm72$  & $1915\pm206$ & $551\pm10$ & $300\pm10$ & 0.05       & $1.80\pm0.22$ & 1  
\enddata
\tablecomments{Col.~(1): Object name(s).  Col.~(2): SED classification type (FS 
= Flat Spectrum); see \S 3.3 for details.  Col.~(3): Stellar spectral type from 
the literature (see \S 4.2).  Col.~(4): 350\,$\mu$m flux density.  Col.~(5): 
450\,$\mu$m flux density.  Col.~(6): 850\,$\mu$m flux density.  Col.~(7): 
1.3\,mm flux density.  Col.~(8): Disk mass (see \S 3.1).  Col.~(9): 
Submillimeter continuum slope (see \S 3.2).  Col.~(10): Notes on individual 
sources as follows: (1) The 1.3\,mm flux densities are from \citet{am94}.  The 
1\,$\sigma$ rms uncertainties were typically assumed to be $\sim$10\,mJy.  (2) 
800\,$\mu$m flux densities from \citet{jensen96} are listed in Col.~(6).  (3) 
The 1.3\,mm flux densities are from \citet{nurnberger98}.  (4) The 1.3\,mm flux 
densities are from \citet{motte98}.  (5) The 1.3\,mm flux densities for WL 20 
(S) and WSB 60 are from the interferometric survey of \citet{andrews07}.  (6) 
The 1.3\,mm flux densities are from \citet{stanke06}.  All flux densities are 
measured in units of mJy.  Upper limits are taken at the 3\,$\sigma$ confidence 
level.  Quoted errors are the 1\,$\sigma$ rms noise levels and do not include 
systematic errors in the absolute flux calibration ($\sim$25\%\ at 350 and 
450\,$\mu$m, $\sim$10\%\ at 850\,$\mu$m, and $\sim$20\%\ at 1.3\,mm).}
\end{deluxetable}

\clearpage

\begin{deluxetable}{lcccccc|lcccccc}
\tablecolumns{14}
\tabletypesize{\scriptsize}
\tablewidth{0pc}
\tablecaption{Results of SED Fits \label{sedfits_table}}
\tablehead{
\colhead{} & \colhead{$T_1$} & \colhead{} & \colhead{$M_d$} & \colhead{} & \colhead{} & \colhead{} & \colhead{} & \colhead{$T_1$} & \colhead{} & \colhead{$M_d$} & \colhead{} & \colhead{} & \colhead{} \\
\colhead{Object} & \colhead{[K]} & \colhead{$q$} & \colhead{[$10^{-3}$ M$_{\odot}$]} & \colhead{$\tilde{\chi}^2$} & \colhead{dof} & \colhead{Notes} & \colhead{Object} & \colhead{[K]} & \colhead{$q$} & \colhead{[$10^{-3}$ M$_{\odot}$]} & \colhead{$\tilde{\chi}^2$} & \colhead{dof} & \colhead{Notes} \\
\colhead{(1)} & \colhead{(2)} & \colhead{(3)} & \colhead{(4)} & \colhead{(5)} & \colhead{(6)} & \colhead{(7)} & \colhead{(1)} & \colhead{(2)} & \colhead{(3)} & \colhead{(4)} & \colhead{(5)} & \colhead{(6)} & \colhead{(7)}}
\startdata
AS 205   & 304 & 0.55 & $33\pm5$    & 2.6 & 7 & 1,2,3   & IRS 49   & 142 & 0.58 & $2.9\pm0.6$ & 0.5 & 3 & 1,2,5 \\
SR 4     & 181 & 0.47 & $3.8\pm0.5$ & 0.6 & 4 & 1,2,4   & WSB 52   & 162 & 0.52 & $7\pm3$     & 0.1 & 3 & 1,2,5,7 \\
GSS 26   & 157 & 0.63 & $45\pm11$   & 3.8 & 4 & 1,2,4   & SR 9     & 150 & 0.56 & $1.9\pm0.9$ & 8.7 & 2 & 1,2,5 \\
EL 20    & 157 & 0.69 & $33\pm9$    & 3.0 & 3 & 1,2,5   & WSB 60   & 99  & 0.57 & $32\pm8$    & 4.0 & 6 & 1,2,6 \\
DoAr 24E & 218 & 0.63 & $8\pm1$     & 0.3 & 5 & 1,2,5,6 & SR 13    & 189 & 0.63 & $14\pm3$    & 1.0 & 6 & 1,2,3 \\
EL 24    & 229 & 0.65 & $162\pm62$  & 1.5 & 8 & 1,3,5,6 & DoAr 44  & 169 & 0.54 & $10\pm2$    & 1.2 & 3 & 1,2,3 \\
EL 26    & 116 & 0.70 & $5\pm3$     & 2.2 & 2 & 1,2,5   & RNO 90   & 215 & 0.59 & $4.7\pm0.7$ & 5.7 & 5 & 1,2,3 \\
IRS 39   & 104 & 0.69 & $14\pm5$    & 2.0 & 4 & 1,2,3,6 & Wa Oph 6 & 157 & 0.65 & $81\pm23$   & 1.5 & 5 & 3,8 \\
YLW 16C  & 120 & 0.63 & $26\pm10$   & 0.2 & 2 & 1,2,5   & AS 209   & 230 & 0.63 & $52\pm15$   & 1.0 & 4 & 3 \\
GY 292   & 143 & 0.63 & $7\pm3$     & 2.3 & 2 & 1,2,6   &          &     &      &             &     &   &  \\
\enddata
\tablecomments{Col.~(1): Object name.  Col.~(2): Best-fit value of the 
temperature at 1\,AU in K (typical errors are $\pm$ a few K).  Col.~(3): 
Best-fit value of the radial power-law index of the temperature profile 
(typical errors are $\pm$0.02).  Col.~(4): Best-fit value and uncertainty of 
the disk mass.  Col.~(5): Reduced $\chi^2$ statistic.  Col.~(6): Number of 
degrees of freedom in the fit (i.e., number of data points minus number of 
fitted parameters[=3]).  Col.~(7): Notes on literature sources for the SED 
data as follows: (1) {\it Spitzer} Space Telescope IRAC flux densities (3.6, 
4.5, 5.8, 8\,$\mu$m) from L.~Allen (2007, private communication).  (2) {\it 
Spitzer} Space Telescope MIPS flux density (24\,$\mu$m) from L.~Allen (2007, 
private communication).  (3) {\it IRAS} flux densities (12, 25, 60\,$\mu$m) 
from \citet{weaver92}.  (4)  Mid-infrared photometry (10, 20\,$\mu$m) from 
\citet{wilking89}.  (5) Mid-infrared photometry (10, 12\,$\mu$m) from 
\citet{barsony05}.  (6) {\it ISO} photometry (6.7 and 14.3\,$\mu$m) from 
\citet{bontemps01}.  (7) Mid-infrared photometry (10, 20\,$\mu$m) from 
\citet{greene94}.  (8) {\it Spitzer} IRAC and MIPS photometry from 
\citet{padgett06}.}
\end{deluxetable}

\clearpage
                                                                                
\begin{deluxetable}{lcccc}
\tablecolumns{5}
\tabletypesize{\scriptsize}
\tablewidth{0pc}
\tablecaption{Statistical Comparison of $\rho$ Oph and Taurus Submillimeter Properties\label{stats_compare}}
\tablehead{
\colhead{} & & & \colhead{$P(M_d)$} & \colhead{$P(\alpha)$} \\
\colhead{Samples} & & & \colhead{[\%]} & \colhead{[\%]}}
\startdata
\it{Regional Comparison} \\
Tau vs. Oph (all)   & & & $35-47$       & $9-62$      \\
Tau I vs. Oph I     & & & $20-72$       & $54-55$     \\
Tau II vs. Oph II   & & & $7-10$        & $3-65$      \\
Tau III vs. Oph III & & & $9-34$        & \nodata     \\
\hline
\it{Evolutionary Comparison} \\
Oph I vs. Oph II    & & & $\ge 99.75$   & $92-95$     \\
Oph II vs. Oph III  & & & $\ge 99.9999$ & \nodata     \\
Tau I vs. Tau II    & & & $\ge 99.97$   & $\ge 99.91$ \\
Tau II vs. Tau III  & & & $\ge 99.9999$ & \nodata     \\
\enddata
\tablecomments{Results of survival analysis two-sample statistical tests 
comparing various subsamples of the submillimeter properties of Taurus (Paper 
I) and $\rho$ Oph YSOs.  The listed values $P$ correspond to the probabilities 
that the two subsamples are drawn from {\it different} parent distributions.  
The probability ranges are from different statistical tests.}
\end{deluxetable}

\clearpage

\begin{deluxetable}{lcccc}
\tablecolumns{5}
\tabletypesize{\scriptsize}
\tablewidth{0pc}
\tablecaption{Median Submillimeter Properties of YSOs \label{benchmark_table}}
\tablehead{
\colhead{} & \colhead{$\langle M_d \rangle$} & \colhead{$\langle M_d/M_{\ast} \rangle$} & \colhead{} \\
\colhead{Sample} & \colhead{[M$_{\odot}$]} & \colhead{[\%]} & \colhead{$\langle \alpha \rangle$}}
\startdata
Class I                         & 0.015 	     & 3.8     & 2.3     \\
Class II                        & 0.005		     & 0.9     & 1.9     \\
transition\tablenotemark{a}     & 0.0009	     & 0.3     & \nodata \\
Class III                       & $< 5\times10^{-5}$ & \nodata & \nodata 
\enddata
\tablenotetext{a}{Transition objects are defined here to be Class III sources 
(i.e., $n > 1.6$) that have submillimeter detections.  There is 1 such object 
in $\rho$ Oph (WL 5) and 7 others in Taurus (CIDA-8, CoKu Tau/4, FW Tau, FY 
Tau, HQ Tau, LkH$\alpha$ 332/G1, and V807 Tau).  The submillimeter detection of 
WL 5 may be anomalous: \citet{motte98} showed that the original quoted 1.3\,mm 
detection from earlier work \citep{am94} was actually background emission.  The 
Class III sample median $M_d$ is based on the stacked average limit derived in 
\S 4.1.}
\end{deluxetable}

\clearpage

\clearpage

\clearpage

\clearpage

\clearpage

\clearpage
                                                                                
\clearpage
                                                                                
\clearpage
                                                                                
\clearpage

\clearpage

\clearpage
                                                                                
\end{document}